\documentclass[aps,twocolumn,pre,superscriptaddress,amsmath,runinaddress]{revtex4}
\usepackage{graphicx}
\usepackage{subfigure}
\usepackage{psfrag}
\usepackage{bm}

\begin{document}


\title{Probability of metastable configurations in spherical three-dimensional Yukawa crystals}


\author{H. K\"{a}hlert} \author{P. Ludwig} \author{H. Baumgartner} \author{M. Bonitz}
\affiliation{Institut f\"{u}r Theoretische Physik und Astrophysik, Christian-Albrechts Universit\"{a}t zu Kiel, 24098 Kiel, Germany}
\author{A. Piel} \author{D. Block}
\affiliation{Institut f\"{u}r Experimentelle und Angewandte Physik, Christian-Albrechts Universit\"{a}t zu Kiel, 24098 Kiel, Germany}
 \author{A. Melzer}
\affiliation{Institut f\"{u}r Physik, Ernst-Moritz-Arndt Universit\"{a}t, 17487 Greifswald, Germany}


\date{\today}

\begin{abstract}
Recently the occurrence probabilities of ground- and metastable states of three-dimensional Yukawa clusters with 27 and 31 particles have been analyzed in dusty plasma experiments~[Block et al., Physics of Plasmas 15, 040701 (2008)]. There it was found that, in many cases, the ground state appeared substantially less frequently than excited states. Here we analyze this question theoretically by means of molecular dynamics (MD) and Monte Carlo simulations and an analytical method based on the canonical partition function. We confirm that metastable states can occur with a significantly higher probability than the ground state. The results strongly depend on the screening parameter of the Yukawa interaction and the damping coefficient used in the MD simulations. The analytical method allows one to gain insight into the mechanisms being responsible for the occurrence probabilities of metastable states in strongly correlated finite systems.
\end{abstract}


\maketitle

\section{Introduction}
Finite strongly coupled systems of charged particles in external traps are of high interest in many fields. Examples include ion crystals~\cite{Wineland1987,drewsen1998}, quantum dots~\cite{Alex2001} and dusty plasma crystals~\cite{Arp2004,bonitz2008pop}. Dusty plasmas allow for an easy realization of strong coupling in laboratory experiments. They typically consist of $\mu m$ sized particles in an rf discharge. Due to their high mass their motion occurs on a macroscopic timescale which makes them an ideal system for studying dynamical properties in the strong coupling limit. In the case of an isotropic parabolic confinement and (screened) Coulomb interaction the ground states are nested spherical shells (3D) or concentric rings (2D).

For classical systems the ground states are found by minimizing the potential energy $U$ with respect to all particle positions. This can be a difficult task since in general $U$ has many minima which may be energetically very close to each other, particularly in 3D. To find the lowest energy configuration one has to avoid trapping in a metastable state, which can be a serious problem for numerical computations. A detailed analysis of the ground states of 3D Coulomb clusters was presented in~\cite{Patrick2005, Arp2005}. The ground states of small spherical Yukawa clusters for a wide range of the screening parameter can be found in~\cite{Henning}. Besides the ground state also metastable states were found in the simulations~\cite{Patrick2005, Arp2005, Apolinario}. Furthermore a fine structure was observed, i.e. states with the same number of particles on each shell but with a different arrangement on the same shell~\cite{Patrick2005}.

Coulomb or Yukawa balls have been produced in dusty plasma experiments~\cite{Arp2004}. They are well explained by a simple model of harmonically confined particles interacting by a Yukawa potential for $N=100\dots 500$~\cite{bonitz2006}. Recently metastable states of Yukawa balls have been investigated in~\cite{block2008} for small particle numbers  $N=27$ and $N=31$. It was found that often metastable states occurred with a higher probability than the ground state. This was confirmed by MD simulations but no theoretical explanation was given. This is the goal of the present paper. We apply Monte Carlo simulations (MC) as well as extensive molecular dynamics (MD) simulations with a broader parameter range than before, confirming the main results of~\cite{block2008}. For a theoretical explanation we apply an analytical method based on the classical canonical partition function~\cite{baletto2005}.

This paper is organized as follows. In Sec.~\ref{sec:model} we present the Hamiltonian and explain our simulation methods. Results of the MD simulations are given in Sec.~\ref{sec:MDresults}. In Sec.~\ref{sec:analyt} we introduce an analytical method for the probabilities of stationary states in thermodynamic equilibrium. The results are compared to MC simulations. Section~\ref{sec:comp} compares the theoretical results with the experiments. The last section summarizes our findings and discusses the applicability range of our models.

\section{Model and simulation idea}\label{sec:model}
\subsection{\label{ssec:model}Hamiltonian}
The system of $N$ identical particles with charge $Q$ and mass $m$ in an isotropic, parabolic confinement 
\begin{equation}\label{eqn:Vext}
V_{ext}(r)=\frac{m}{2}\omega_0^2 {r}^2
\end{equation}
($r=|\bm{r}|$) is described by the Hamiltonian
\begin{equation}\label{eqn:Hamiltonian}
  H=\sum_{i=1}^{N}\left\{\frac{p_ i ^2}{2m}+V_{ext}({r}_i)\right\} + \sum_{i>j}^N V(|\bm{r} _i-\bm{r} _j|).
\end{equation}
The interaction is assumed to be a shielded Coulomb-potential of the form
\begin{equation}\label{eqn:Yukawa}
  V(r)=\frac{Q^2}{r}e^{-\kappa r},
\end{equation}
where the range of the potential is controlled by the screening parameter $\kappa$. Despite its simplicity this model is of relevance for many systems, such as colloids, and has proven to accurately describe the spherical dust crystals (Yukawa balls) observed in experiments~\cite{bonitz2006}. In dusty plasmas $\kappa$ is given by the inverse Debye screening length. In the following we will treat it as a free parameter and focus on the general behavior of the model~(\ref{eqn:Hamiltonian}). Using, as a particular example, typical dusty plasma parameters will allow us to make comparisons with the experimental observations of Ref.~\cite{block2008}.

Results will be given in units of the distance $r_0=({2Q^2}/{m\omega_0^2})^{1/3}$ and the corresponding Coulomb energy $E_0={Q^2}/{r_0}$. Frequencies and forces are given in units of $\omega_0$ and $m\omega_0^2 r_0$, respectively.

The ground (metastable) states are the global (local) minima of the potential energy $U$,
\begin{equation}
U(\bm{r} _1,\dots,\bm{r} _N)=\sum_{i=1}^{N}V_{ext}(r_i) + \sum_{i>j}^N V(|\bm{r}_i-\bm{r} _j|).
\end{equation}
In both cases the total force on all particles vanishes and the system is in a stable configuration, i.e. stable against small perturbations.

\subsection{\label{ssec:MC}Monte Carlo (MC)}
The MC simulations use the standard Metropolis algorithm~\cite{computational} with the Hamiltonian~(\ref{eqn:Hamiltonian}), but without the kinetic energy part. Starting from the classical ground state at $T=0$ the system is given a finite temperature. For a fixed temperature we performed $10^7$ MC steps and determined the configuration every $10^4th$ step. The temperature is then increased and the same procedure repeated. Ergodicity of the procedure was checked by using different initial configurations. Following this method we calculate the probability as a function of $T$ from the number of occurrences of the different states.

\subsection{\label{ssec:MD}Molecular dynamics (MD)}
 In the MD simulations we follow a different approach. Here we solve the equations of motion for particles in a parabolic trap interacting through the Yukawa potential~(\ref{eqn:Yukawa}) but include an additional damping term to simulate the annealing process the way it occurs in the experiment, as explained in~\cite{block2008}. This is different from the MC simulations where the particles are in contact with a heat bath and maintain a constant temperature. This also differs from the MD simulations in~\cite{block2008} which were also performed at finite temperature. Here, we perform substantially larger simulations and systematically scan a broad parameter range. For the $i$-th particle the equation of motion we solve is
\begin{equation}\label{eqn:EOM}
  m\ddot{\bm{r}}_i=-\nabla_i U(\bm{r} _1,...,\bm{r} _N)-\nu m \dot{\bm{r}}_i,
\end{equation}
where $\nu$ is the collision frequency which will be given in units of $\omega_0$. In dusty plasmas friction is mainly due to the neutral gas.

The simulation is initialized with random particle positions and velocities in a square box. To stop the simulation and determine the configuration we use two similar, but not equivalent conditions:
\begin{enumerate}
 \item[(A)] The particles' mean kinetic energy drops below a value $\left< E_{kin}^{min}\right>$ of typically $10^{-6}-10^{-8}$.
 \item[(B)] The force on each particle due to the confinement and the other particles decreases below $10^{-4}$.
\end{enumerate}
It is tempting to define (A) as a proper condition but we will show that (B) has to be used, although they look equivalent at first glance. The difference lies in the definition of a stable configuration. If the particles lose their initial kinetic energy before they have reached a local minimum the simulation could be stopped before the particle motion has effectively ended. This problem can be circumvented by condition (B) which makes direct use of the definition of a stable state, namely that the force on each particle due to $U$ vanishes.

The screening parameter, the friction coefficient as well as the lower limit for the mean kinetic energy are varied. For each parameter setting the simulation is repeated $3000-5000$ times to obtain accurate statistics. We consider systems with 31 and 27 particles as was done in the experiment. As another example we used a cluster with 40 particles because here the ground state shell configuration abruptly changes from (34,6) to (32,8) at $\kappa=0.415$ as the screening parameter is increased - without the configuration (33,7) ever being the ground state~\cite{Henning}. This gives rise to the question of how often this configuration can actually occur in experiments.

\section{\label{sec:MDresults}MD simulation results}
In this section we present the results of our first-principle MD simulations. The main parameters determining the occurrence frequencies of different metastable states for a given $N$ are the screening parameter $\kappa$ and the friction $\nu$. We therefore discuss the dependence on $\kappa$ and $\nu$ in detail. As an example of particular interest we will consider the parameter values of dusty plasma experiments which are in the range of $\kappa\approx0.4\dots 1.0$~\cite{block2008}. This case will be dealt with in Section~\ref{sec:comp}. 

We first discuss the effect of the damping rate on the occurrence probabilities. It will turn out that with a properly chosen rate we can produce very general results for different screening lengths which do not depend on the exact chosen damping coefficient and hold for any rate in the overdamped limit. The effect of screening will then be examined in the following section.

\subsection{Effect of friction}\label{ssec:friction}
\begin{figure}
 \includegraphics[width=0.45\textwidth]{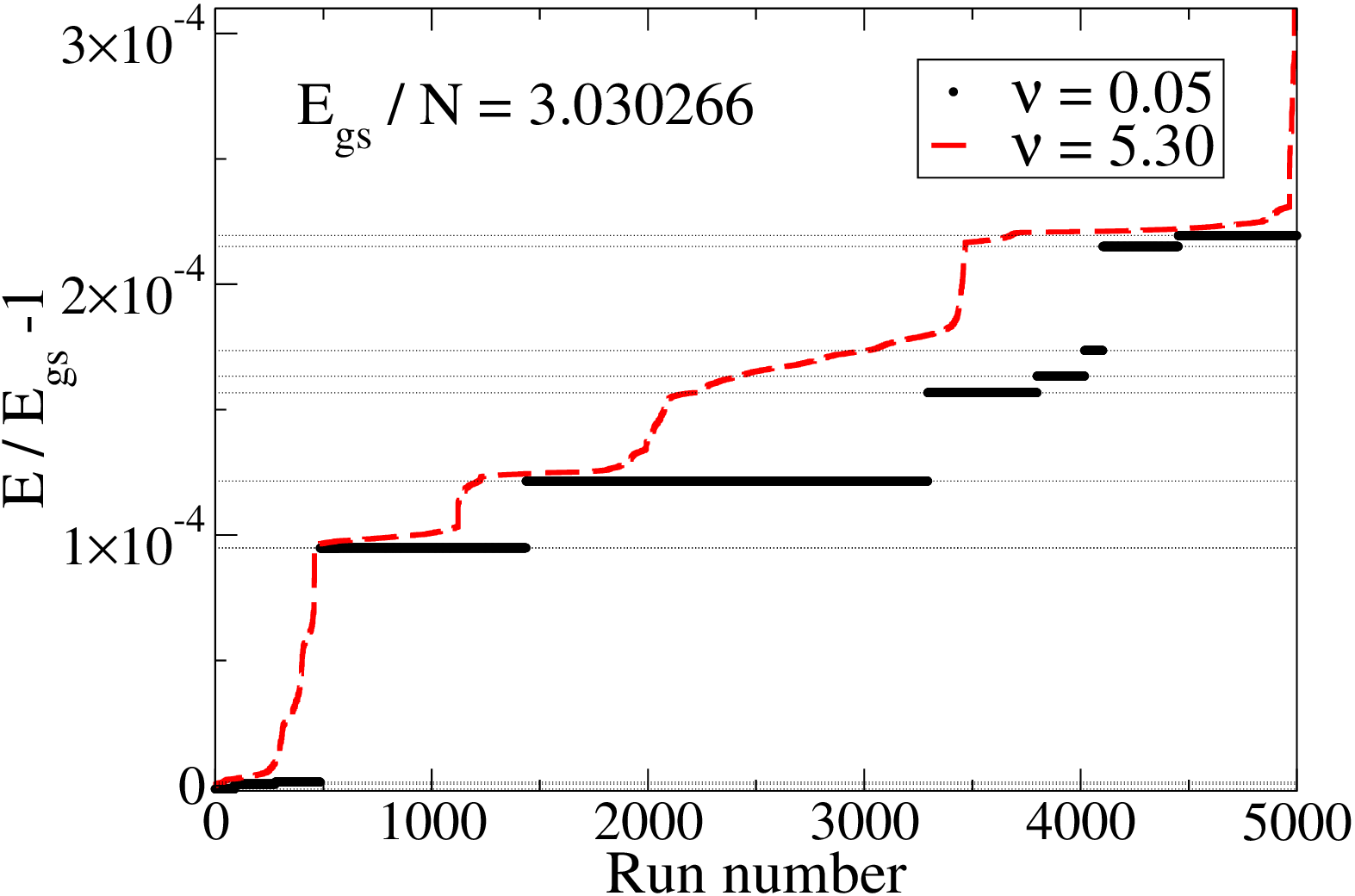}
 \caption{(Color online) Stationary states observed in the MD-simulations for $N=31,\,\kappa=1.4$ and $\left< E_{kin}^{min}\right>=10^{-8}$. The runs are sorted by the energy or the stationary state, see also table~\ref{tab:fricstates}. For slow cooling (black bars, $\nu=0.05$) one can clearly see distinct states which correspond to the horizontal lines. The length of the bold lines is proportional to the occurrence probabilities. In the case of strong friction (red, dashed line, $\nu=5.3$) the particles often lose their kinetic energy before they can settle into the equilibrium positions and the fine structure (different states with same shell configuration) cannot be resolved.\label{fig:fricruns}}
 \end{figure}

A typical simulation result is shown in Fig.~\ref{fig:fricruns}. For slow cooling ($\nu=0.05$) the particles are not hindered by friction and can move according to the interparticle- and confinement forces. They continuously lose kinetic energy until they are trapped in a local minimum of the potential energy $U$. Here they are further being damped until the simulation is stopped. It is interesting to see that there exist more metastable states than different shell occupations, as was first observed in~\cite{Patrick2005}, see also~\cite{Apolinario}. Details are given in Table~\ref{tab:fricstates}.

In the case of strong damping ($\nu=5.3$) the situation is different. Here the particles are readily slowed down after the initialization process in the box. Their motion is strongly affected by friction and interrupted even before they may be trapped in a local minimum. If condition~(A) is used to stop the simulation it is not clear if the particles are in a stable state. The reason is that due to the rapid damping they can be sufficiently slowed even though they are not in a potential minimum but on a descending path and would reach the stable configuration at a later time.
 \begin{table}
 \caption{Energy difference between metastable states and the ground state (the ground state and its energy is given by italic numbers) as seen in Fig.~\ref{fig:fricruns}. States with the same shell configuration but different energy differ only by the arrangement of the particles on the same shell (fine structure).}
 \begin{ruledtabular}\label{tab:fricstates}
 \begin{tabular}{cc|cc}
 $\Delta E/N$ & config. & $\Delta E/N$ & config.\\
 \hline
${\textsl{3.030266}}$ & {\textsl{(27,4)}}& $0.000479$ & (26,5)\\
$0.000006$ & (27,4) & $0.000499$ & (26,5)\\
$0.000009$ & (27,4) & $0.000530$ & (26,5)\\
$0.000291$ & (26,5) & $0.000656$ & (25,6)\\
$0.000372$ & (26,5) & $0.000669$ & (25,6)
 \end{tabular}
 \end{ruledtabular}
 \end{table}
 \begin{figure}
 \centering
 \includegraphics*[width=0.42\textwidth]{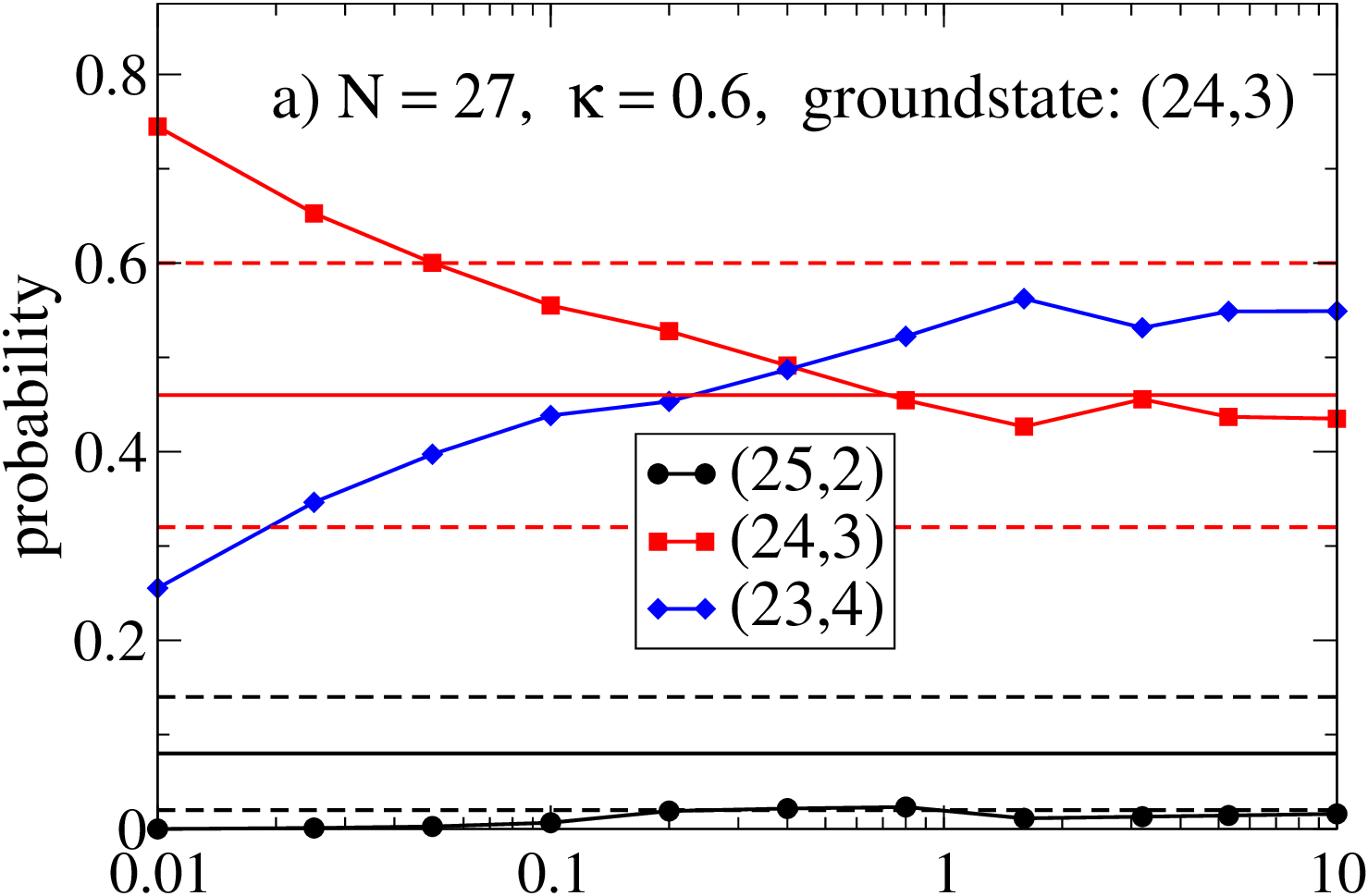}\vspace{0.13cm}
 \includegraphics*[width=0.42\textwidth]{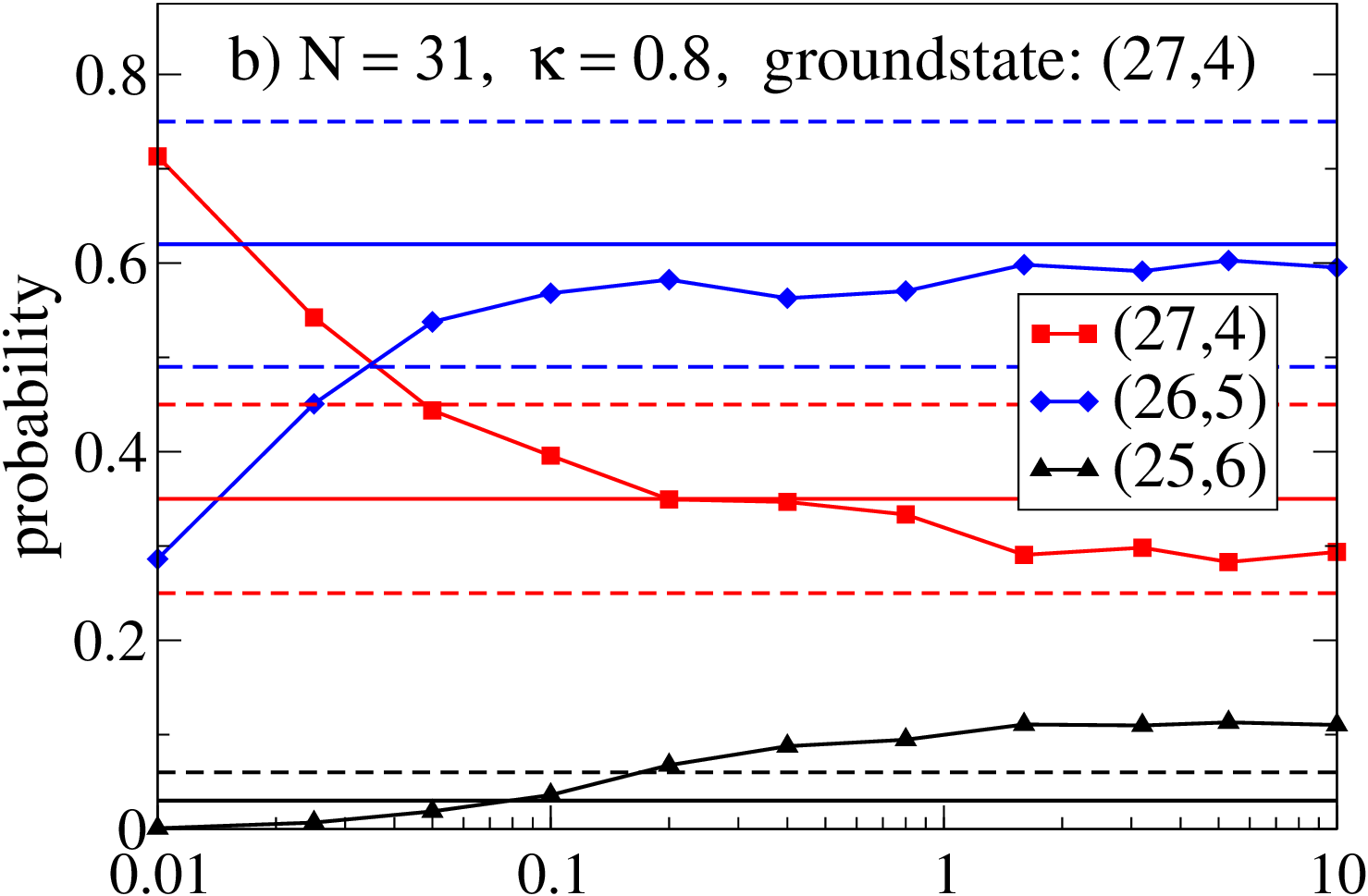}\vspace{0.13cm}
 \includegraphics*[width=0.42\textwidth]{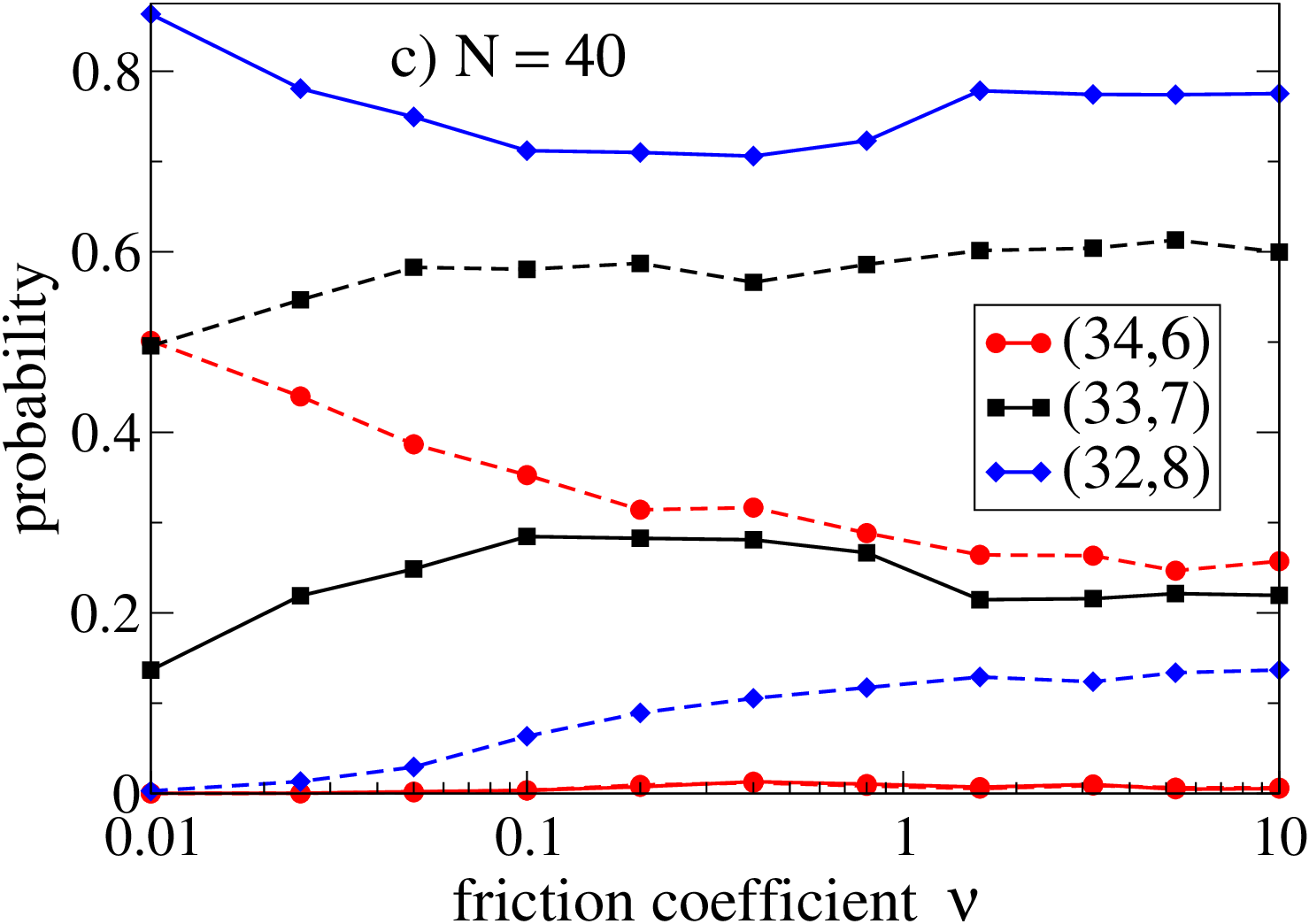}
 \caption{(Color online) Effect of friction on the occurrence probabilities obtained with condition~(B) for three different numbers of particles. In a) and b) horizontal solid and dashed lines indicate experimental mean and standard deviation, respectively~\cite{block2008}. For $N=27$ the experimental values for the clusters (23,4) and (24,3) are the same. In c) solid lines indicate Yukawa interaction with $\kappa=1.0$ [ground state (32,8)] whereas dashed lines show results for Coulomb interaction [ground state (34,6)]. In all cases slow cooling favors the ground states over metastable states.\label{fig:friction}}
 \end{figure}

Fig.~\ref{fig:friction} shows the influence of friction on the occurrence probabilities in more detail. For fixed screening the probability of finding the ground state configuration increases when the friction coefficient is decreased. Here the particles are cooled down more slowly and it is more likely that they reach the system's true ground state. During the cooling process they still have a sufficiently high kinetic energy and time to escape from a local minimum until the force on each particle vanishes.
In the case of strong friction the particles can fall into a nearby minimum and leaving it becomes more difficult due to the rapid loss of kinetic energy. The typical simulation time until the forces are small enough is longer than for intermediate friction strength. Once cooled down the particles are pushed along the gradient of the potential energy surface until they reach a stable state. Thus the results can depend on how far the system's temperature is decreased. One can see that for $\nu>2$, i.e. in the overdamped regime, the probabilities have practically saturated. For fast cooling, i.e. large friction, metastable states can occur with a comparable or even higher probability than the ground state.

The $N=40$ cluster shows a qualitatively different behavior compared to the $N=27,\,31$ clusters. In the case of $\kappa=1.0$ the lines corresponding to different configurations do not intersect and the ground state is the most probable state regardless of the damping coefficient. In contrast, in the Coulomb limit, $\kappa=0$, the most probable state is a always a metastable state, except for very small friction, $\nu \le 0.01$.

Dusty plasma experiments are performed in the overdamped regime, i.e. here $\nu$ is of the order of $3-6$~\cite{block2008}. Since in this limit the probabilities depend only very weakly on the damping rate the results presented in the next section for $\nu=3.2$ should hold for any such damping coefficient. Even though this was shown only for a few examples we believe that this also holds for other particle numbers and screening lengths.

 \begin{figure}
 \centering
{\includegraphics*[width=0.42\textwidth]{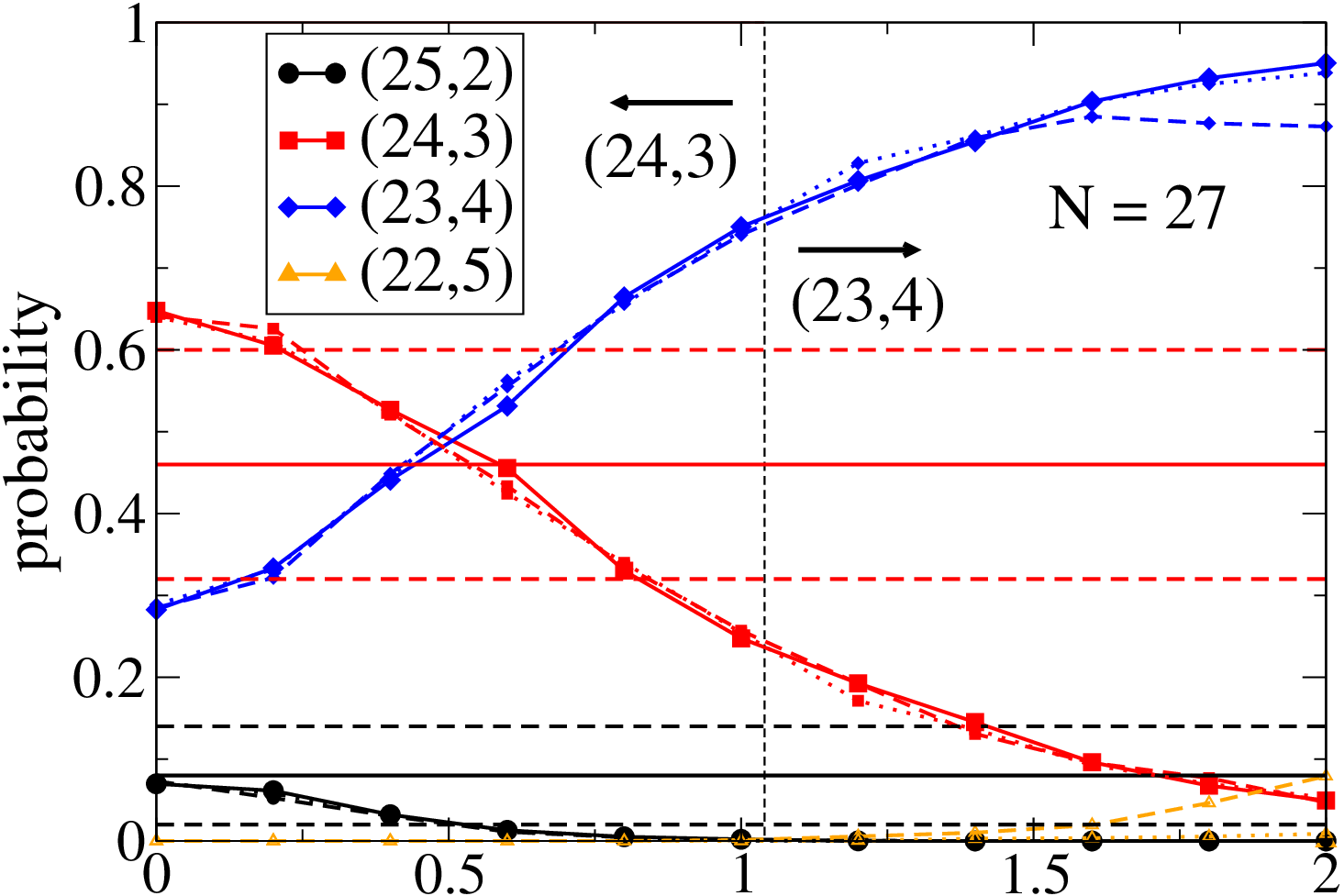}}\vspace{0.13cm}
{\includegraphics*[width=0.42\textwidth]{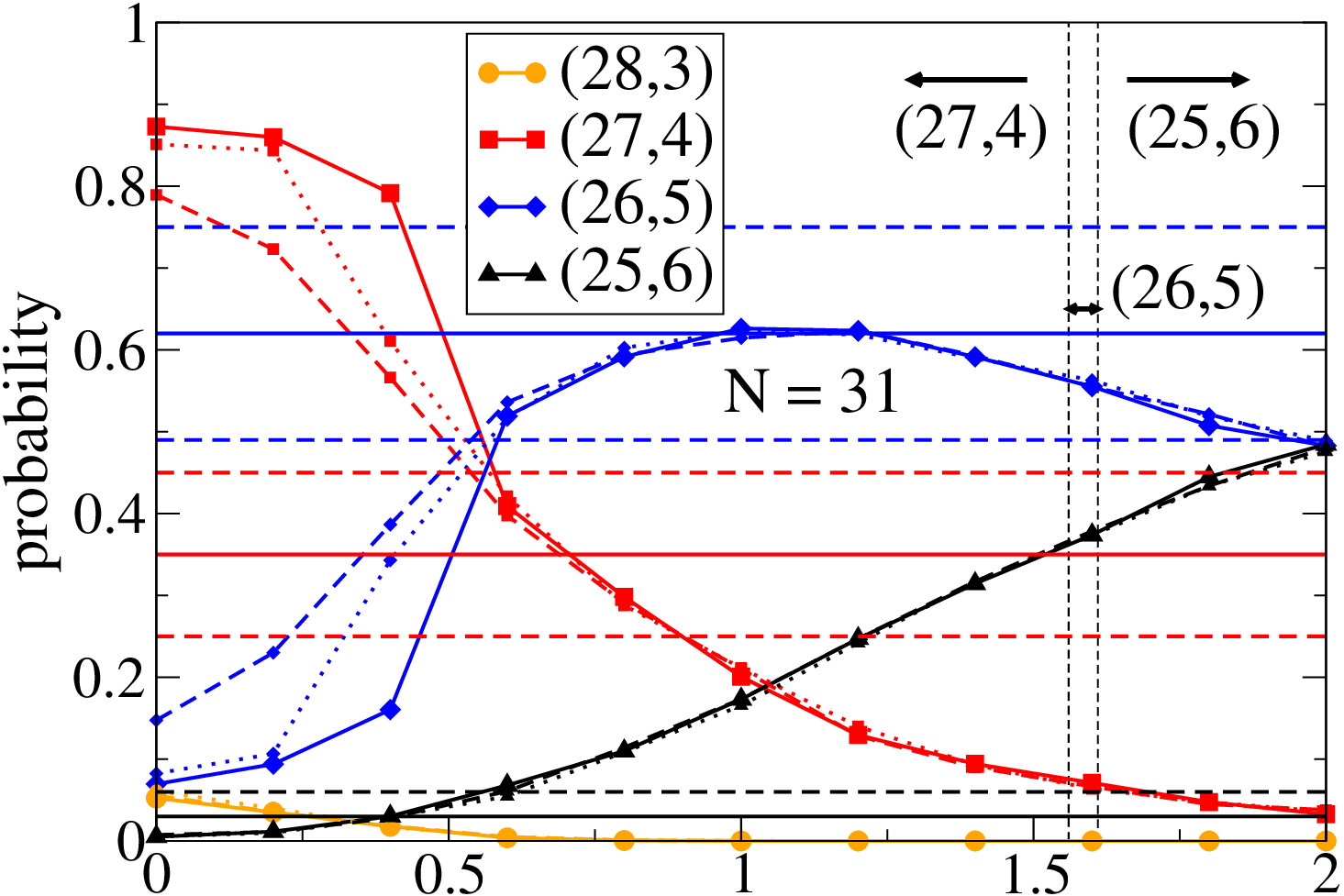}}\vspace{0.13cm}
{\includegraphics*[width=0.42\textwidth]{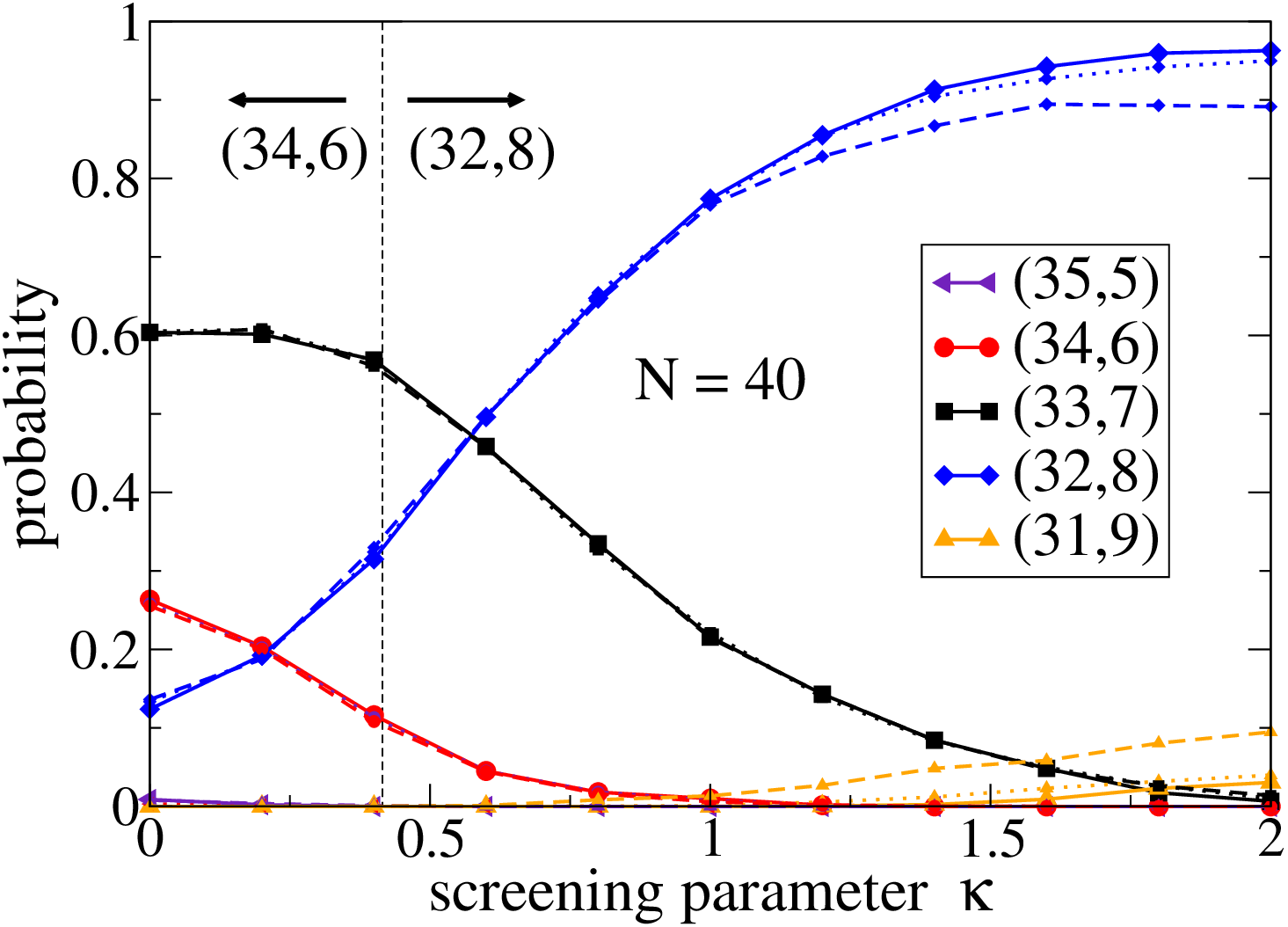}}
 \caption{(Color online) Effect of screening for $\nu=3.2$. Solid lines show the results obtained with condition~(B) while dotted and dashed lines indicate use of condition~(A) with $\left< E_{\text{kin}}^{\text{min}}\right>=10^{-8},\,10^{-7}$, respectively. Arrows show the ground state configuration to the left or right from the vertical line. Where available horizontal solid and dashed lines indicate experimental mean and standard deviation~\cite{block2008}. For the $N=27$ cluster the experimental values for the configurations (23,4) and (24,3) are the same.\label{fig:kappa}}
 \end{figure}

\subsection{Effect of screening}\label{ssec:screening}
The screening dependence of the ground state shell configurations of spherical Yukawa clusters in the absence of damping has been analyzed in Ref. \cite{bonitz2006}. The general trend is that increased screening favors ground state configurations with more particles on the inner shell(s). A systematic analysis in a large range of particle numbers and screening parameters \cite{Henning} confirms this trend. Here we extend this analysis to spherical crystals in the presence of damping and also consider the screening dependence of the occurrence probability of metastable states.

For a fixed friction coefficient in the overdamped limit the effect of screening is shown in Fig.~\ref{fig:kappa}. The different ground state configurations are indicated by the numbers with arrows in the figures. As in the undamped case, at some finite value of $\kappa$, a configuration with an additional particle on the inner shell becomes the ground state. Consider now the probability to 
observe the ground and metastable states. 
For weak screening the ground states (27,4) and (24,3) are the most probable states in the cases $N=31$ and $N=27$, respectively. At the same time in both cases, the probability of the configuration with one more particle on the inner shell grows with $\kappa$, until it eventually becomes even more probable than the ground state. Note that this occurs much earlier (at a significantly smaller value of $\kappa$) than the ground state change. 
For $N=31$ this trend is observed twice: the probability of the configuration (26,5)
first increases  with $\kappa$ and reaches a maximum around $\kappa\approx 1$. For $\kappa>2$ this configuration becomes less probable than the configuration (25,6), i.e. again a configuration with an additional particle on the inner shell becomes more probable with increased screening.

Different behavior is observed for the $N=40$  cluster where the ground state for weak screening (34,6) is never the most probable state. For large screening, $\kappa\ge 0.6$, the new ground state (32,8) has the highest probability, but this happens only substantially later (for larger $\kappa$) after this state has become the energetically lowest one. This is due to the existence of a third state (33,7) which has the highest probability for $\kappa\le 0.6$ although it is never the energetically lowest one.

Summarizing the above observations we confirm that in spherical Yukawa clusters the ground state is not necessarily the most probable state. Often, a metastable state with more particles on the inner shell is observed substantially (in some cases up to five times) more frequently.
Further, increased screening tends to favor states with more particles on the inner shell. We will give an explanation for this behavior in the next section by using an analytical model for the partition function.

Before doing this we comment on some technical details which are important in the present MD simulations. For certain intervals of the screening parameter the results for the probabilities depend on how far the system is cooled down. Here one state (generally the ground state) is favored over another the smaller $\left< E_{kin}^{min}\right>$ is chosen. This also means increasing the mean simulation time. As discussed before the particles are heavily damped and lose their initial kinetic energy on a short timescale. Their motion is then determined by the shape of the energy surface. Using condition~(B) to terminate the simulation we obtain converged results where the particles have reached a local minimum. Thus if the simulation would be continued the configuration would remain the same.  


\section{Analytical Theory of stationary state probabilities}{\label{sec:analyt}}

\subsection{{\label{ssec:approx}}Harmonic approximation}
The analytical approach to calculating the occurrence probabilities is based on the classical canonical partition function $Z(T,\omega_0,N)$. Instead of the dependence on volume (or density) as in a homogeneous system, here thermodynamic quantities depend on the confining strength $\omega_0$. The partition function can be evaluated analytically in the harmonic approximation, see e.g. Ref.~\cite{baletto2005}. Here the potential energy of a given state is expanded around a local minimum with energy $E_s^0$, where $s$ denotes the ground- or a metastable state. It can be written as
\[
  U_s\approx E_s^0+\frac{1}{2}\sum_{i,j=1}^N\sum_{\alpha,\beta=1}^{3}\left .\frac{\partial^2 U(\bm{r})}{\partial \bm{r}_{i,\alpha} \partial\bm{r}_{j,\beta}}\right|_{\bm{r}=\bm{r}^{0s}} \delta \bm{r}_{i,\alpha}\delta\bm{r}_{j,\beta},
\]
where $\bm{r}^{0s}=(\bm{r}_{1}^{0s},\dots,\bm{r}_{N}^{0s})$ denotes the $3N$-dimensional vector of the particles' equilibrium positions and $\delta\bm{r}_{i,\alpha}=\bm{r}_{i,\alpha} - \bm{r}_{i,\alpha}^0$ the displacement vector. Transforming to normal coordinates $\xi_{s,i}$ this turns into a sum of decoupled harmonic oscillators
\begin{equation}\label{eqn:expansion}
  U_s\approx E_s^0+\frac{1}{2}\sum_{i=1}^{f}m\omega_{s,i}^2\xi_{s,i}^2, \hspace{0.25cm} f=3N-3,
\end{equation}
with eigenfrequencies $\omega_{s,i}$, which are the square roots of the eigenvalues of the Hessian
\[
U_{i,\alpha,j,\beta}=\left .\frac{\partial^2 U(\bm{r})}{\partial \bm{r}_{i,\alpha} \partial\bm{r}_{j,\beta}}\right|_{\bm{r}=\bm{r}^{0s}}.
\]
The expansion~(\ref{eqn:expansion}) includes the particles' three center of mass oscillations in the trap with $\omega=1$ (in units of $\omega_0$). Furthermore we assume that the vibrational and the three rotational modes of the whole system ($\omega=0$) are decoupled, the latter are, therefore, eliminated from the sum (\ref{eqn:expansion}). In the principal axes frame the rotational kinetic energy can then be expressed as
\[
  T_s^{rot}=\sum_{i=1}^3 \frac{L_{s,i}^2}{2I_{s,i}},
\]
with angular momenta $L_{s,i}$ and constant principal moments of inertia $I_{s,i}$. In this approximation the full energy of the state $s$ is, to second order in the displacements,
\begin{eqnarray}\label{eqn:energyapprox}
  E_s=E_s^0+\sum_{i=1}^{f}\left\{\frac{p_{\xi_{s,i}}^2}{2m}+\frac{m}{2}\omega_{s,i}^2\xi_{s,i}^2\right\}+\sum_{i=1}^3 \frac{L_{s,i}^2}{2I_{s,i}}.
\end{eqnarray}
The first term in parentheses denotes the vibrational kinetic energy $T_s^{vib}$.

The harmonic approximation is only applicable for low temperatures (or strong coupling) when the particles oscillate around the equilibrium positions with a small amplitude.

\subsection{{\label{ssec:partition}}Partition function}
The general form of the classical canonical partition function is
\begin{equation}{\label{eqn:classical}}
 Z_s=\frac{n_s}{(2\pi\hbar)^{3N}}\int_{-\infty}^{\infty}dp^{3N} dq^{3N} e^{-\beta H^s(p_i,q_i)}.
\end{equation}
Here it is written for a general Hamiltonian $H^s(p_i,q_i)$ with $3N$ degrees of freedom, generalized coordinates $q_i$ and conjugate momenta $p_i$. Since in our case the energy contributions are independent it can be factorized according to
\begin{equation}\label{eqn:PartitionFunction}
  Z_s=n_sZ_s^{int} Z_s^{vib} Z_s^{rot}
\end{equation}
with the internal partition function
\begin{equation}
Z_s^{int}=e^{-\beta E_s^0}
\end{equation}
and the degeneracy factor $n_s$ calculated as 
\begin{equation}\label{eqn:deg}
n_s=\frac{N!}{\prod_{i=1}^L N_i^s!},
\end{equation}
where $L$ is the number of shells and $N_i^s$ the occupation number of shell $i$ with $\sum_{i=1}^L N_i=N$. The degeneracy factor $n_s$ denotes the number of possibilities to form a configuration with shell occupation $(N_1,N_2,\dots,N_L)$ from distinguishable particles.

$Z_s^{vib}$ is the partition function for $f$ independent harmonic oscillators while $Z_s^{rot}$ is related to the rotational degrees of freedom. The results for our specific case with the energy given by Eq.~(\ref{eqn:energyapprox}) can be found in~\cite{baletto2005} and read
\begin{subequations}\label{eqn:twopartition}
\begin{align}
 Z_s^{vib}(T)&=\left(\frac{k_B T}{\hbar \Omega_{s}}\right)^f,\\
 Z_s^{rot}(T)&=\left(\frac{2\pi k_B T \bar{I} _s}{\hbar^2}\right)^{3/2}.
\end{align}
\end{subequations}
The expressions include the mean geometric eigenfrequency $\Omega_s=(\prod_{i=1}^f \omega_{s,i})^{1/f}$ and the mean moment of inertia $\bar{I} _s=(I_{s,1} I_{s,2} I_{s,3})^{1/3}$.

To obtain the total partition function $Z(T,\omega_0,N)$ the contributions of all $M$ (metastable) states are summed up, i.e.
\[
 Z=\sum_{\sigma=1}^{M} n_{\sigma} Z_{\sigma}.
\]

\subsection{Probability of stationary states}
Collecting the results of subsection~\ref{ssec:partition}, the stationary state probabilities are  given by
\begin{equation}\label{eqn:Probability}
  P_s=\frac{n_s Z_s}{Z}=\frac{n_s Z_s}{\sum_{\sigma=1}^{M} n_{\sigma} Z_{\sigma}}.
 \end{equation}
For our clusters of interest with $27-40$ particles the moments of inertia for different states are equal to a good approximation~(cf.~Table~\ref{tab:inertia} for $N=27$) and can be canceled. Similar behavior is observed for $N=31,\,40$. For low particle numbers, $N\lesssim 10$, they should be included, since here a slight change of the configuration can alter the moment of inertia by a significant amount, but this is not of importance for the present analysis.
 \begin{table}
 \caption{Mean shell radii $R_1,\,R_2$ of first and second shell for states observed in the MD simulations for $N=27$ and $\kappa=0.6$. The relative statistical weight $\tilde{q}_s=(\bar{I}_s/\bar{I}_1)^{3/2}$ caused by different moments of inertia can be neglected in the computation of the probabilities since $\tilde{q}_s\approx 1$ for all states.}
 \begin{ruledtabular}\label{tab:inertia}
 \begin{tabular}{cccccc}
 state $s$& configuration &  $R_2$ & $R_1$ & $\tilde{q}_s$  \\
 \hline
 1 & (24,3) & $1.6175$ & $0.5977$ & $1$\\
 2 & (23,4) & $1.6413$ & $0.6963$ & $1.0009$\\
 3 & (23,4) & $1.6413$ & $0.6957$ & $1.0009$\\
 4 & (25,2) & $1.5935$ & $0.4542$ & $1.0004$\\
 5 & (25,2) & $1.5934$ & $0.4543$ & $1.0004$\\
 \end{tabular}
 \end{ruledtabular}
 \end{table}

Using Eqs.~(\ref{eqn:twopartition}) we obtain from Eq.~(\ref{eqn:Probability})
\begin{equation}\label{eqn:prob}
 P_s\approx \frac{n_s e^{-\beta E_s^0} \Omega_s^{-f}}{\sum_{\sigma=1}^{M}n_{\sigma}  e^{-\beta E_{\sigma}^0} \Omega_{\sigma}^{-f}}.
\end{equation}
To avoid computation of the full partition function [denominator of Eq.~(\ref{eqn:prob})] it is advantageous to compute probability ratios of two states $s$ and $s'$
\begin{eqnarray}\label{eqn:ratioprob}
  \frac{P_s}{P_{s'}}&=&\frac{n_s}{n_{s'}}\left(\frac{\Omega_{s'}}{\Omega_s}\right)^f\left(\frac{\bar{I}_s}{\bar{I}_{s'}}\right)^{3/2}e^{-\beta(E_s^0-E_{s'}^0)}\nonumber\\
&\approx&\frac{n_s}{n_{s'}}\left(\frac{\Omega_{s'}}{\Omega_s}\right)^fe^{-\beta(E_s^0-E_{s'}^0)}.
\end{eqnarray}
Thus the probability ratio of two states depends on three factors: their energy difference $E_s^0-E_{s'}^0$, the ratio of degeneracy factors $n_s/n_{s'}$ and the ratio of mean eigenfrequencies $\Omega_{s'}/\Omega_s$.

The Boltzmann factor $e^{-\beta(E_s^0-E_{s'}^0)}$ gives preference to states with a low energy. For low temperatures it will be the most dominant factor but it becomes less important for higher temperatures when $k_BT\gg E_{s'}^0-E_s^0$ and $e^{-\beta(E_s^0-E_{s'}^0)}\approx 1$.

According to Eq.~(\ref{eqn:deg}) the degeneracy factor assigns a large statistical weight to states with more particles on inner shells. As an example, for $N=27$, we obtain $n_{(25,2)}/n_{(23,4)}=\frac{23!4!}{25!2!}=1/50$. One can see that the configuration with only 2 particles on the inner shell is suppressed due to a lower degeneracy factor contrary to the states with an inner shell consisting of 4 particles, see also Table~\ref{tab:statesN27}. The reason is that there exist more combinatorial possibilities to construct configurations when the difference between the single shell occupation numbers is small. For $N=31$ (Table~\ref{tab:statesN40}) this ratio can be even larger. This shows that (even for low temperatures) this factor can strongly influence the occurrence probabilities.

In the MD simulations we observe several states with the same shell configuration but different energies. Their energy difference can be as large as between states with different configurations (cf. Table.~\ref{tab:fricstates}). In Eq.~(\ref{eqn:Probability}) all states with the same shell configuration are added with the same degeneracy factor.

\begin{figure}[h]
 {\includegraphics*[width=0.42\textwidth]{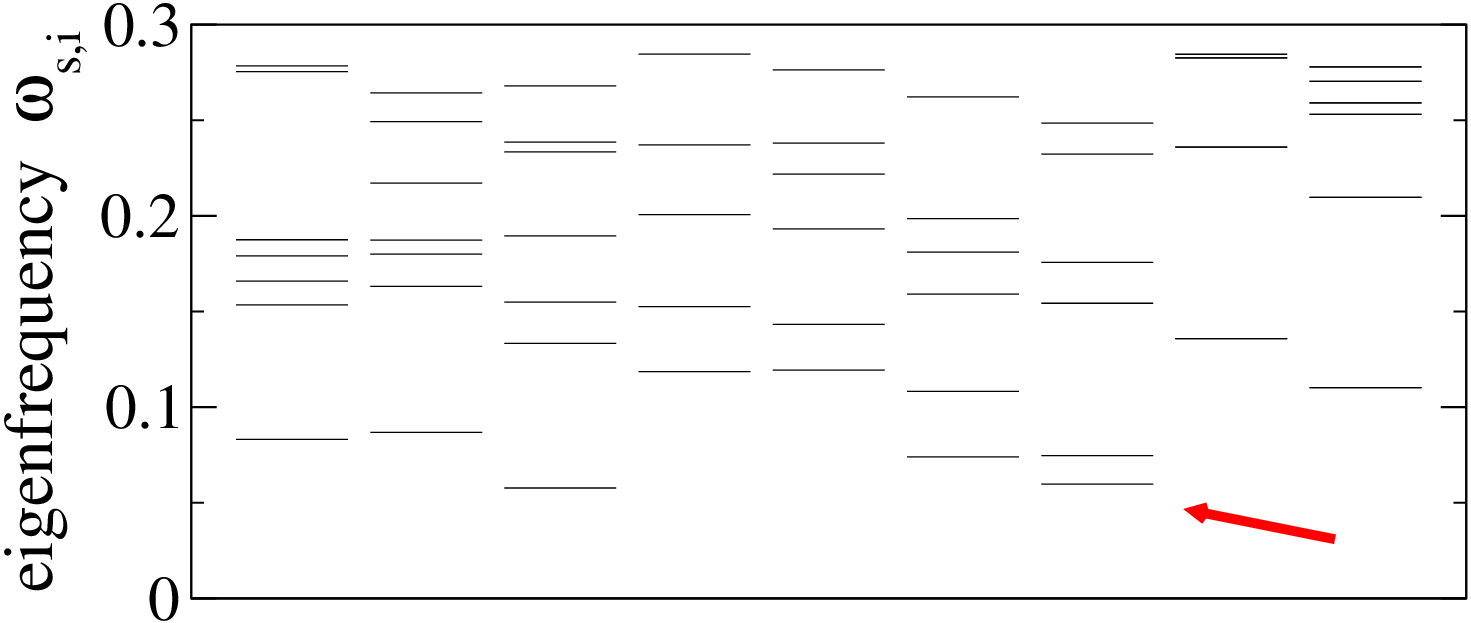}}
 {\includegraphics*[width=0.42\textwidth]{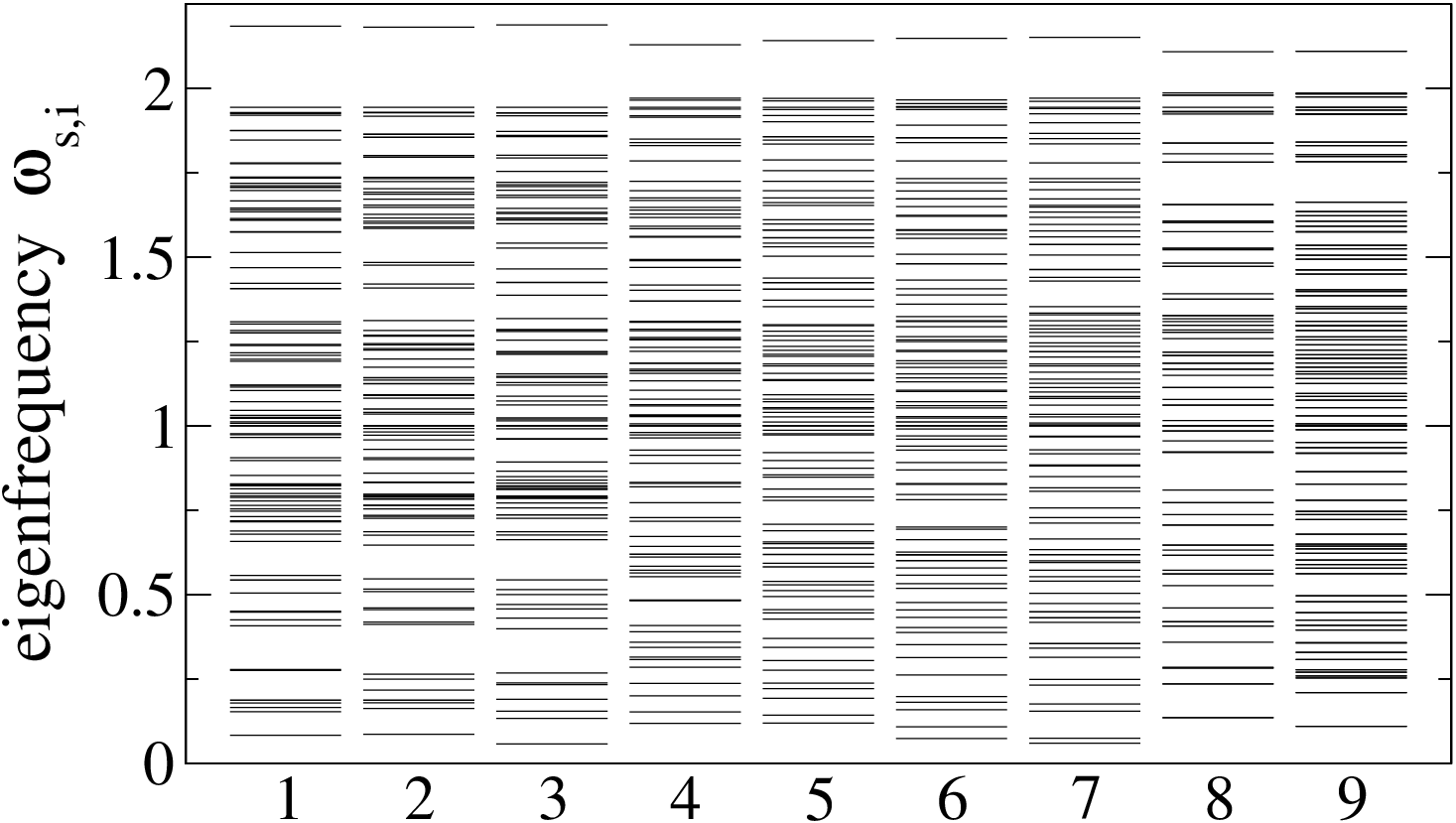}}
 \caption{Spectrum of the eigenfrequencies for the 9 states shown in Table~\ref{tab:statesN31}. The top figure shows the lowest modes in more detail.}{\label{fig:spectrum}}
 \end{figure}

Let us now consider the effect of the mean eigenfrequency, i.e. the effect of the local curvature of the potential energy surface. Written out explicitly, using Eq.~(\ref{eqn:ratioprob}), this factor reads
\begin{equation}
 \left(\frac{\Omega_{s'}}{\Omega_s}\right)^f=\frac{\prod_{i=1}^f \omega_{s',i}}{\prod_{i=1}^f \omega_{s,i}},
\end{equation}
i.e. it is the inverse ratio of the products of the eigenfrequencies. The main contribution here usually arises from the lowest eigenfrequencies. This can be seen in Fig.~\ref{fig:spectrum} showing the spectrum for the states of the cluster with $N=31,\,\kappa=0.8$. State \#7 has two very low eigenfrequencies~[cf. Fig.~\ref{fig:spectrum}, red arrow]  which strongly increase its statistical weight (see also Table~\ref{tab:statesN31}).

\begin{figure}
 {\includegraphics*[width=0.45\textwidth]{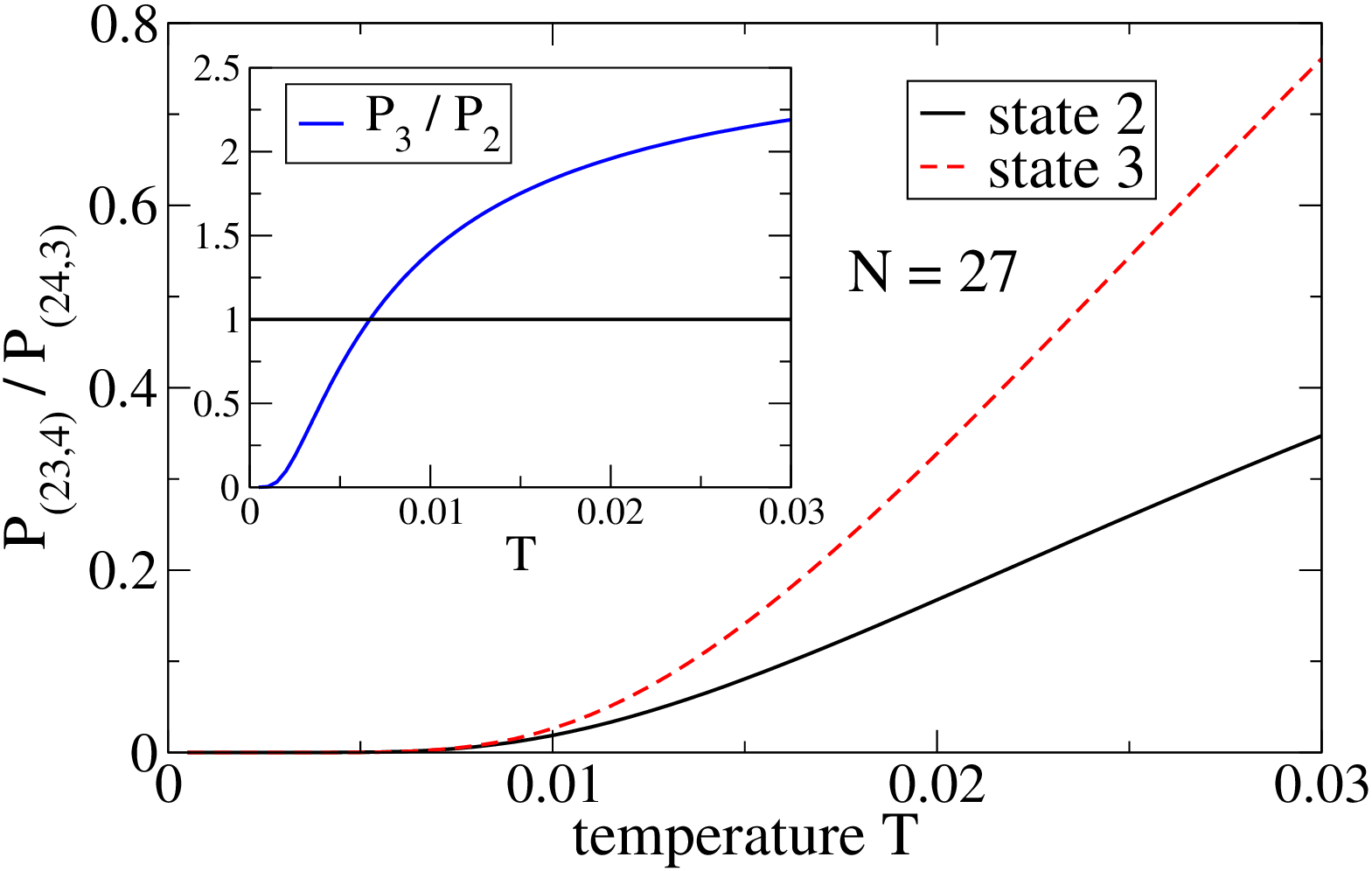}}
 \caption{Probability of the two metastable states with conf. (23,4) compared to the ground state (24,3) for the Yukawa ball with $N=27$. The inset shows the ratio of the probabilities for states 2 and 3 from Table~\ref{tab:statesN27} at low temperatures. Although state 3 has the same configuration and a higher energy the probability of finding state 3 is higher for $T\ge 0.007$ due to the effect of the eigenfrequencies.}{\label{fig:ratio}}
 \end{figure}

For two states with the same shell configuration we have $n_s=n_{s'}$, and the probability ratio is only determined by their energy difference and eigenfrequencies. Even though a state has a higher energy it can have a higher probability provided it has a lower mean eigenfrequency. Fig.~\ref{fig:ratio} shows the effect for $N=27$, for the states listed in Table~\ref{tab:statesN27}.
The physical explanation of the eigenfrequency factor is very simple: states with low eigenfrequencies have a broad (flat) potential energy minimum and thus a larger phase space volume of attraction for the trajectories of $N$ particles. Thus initially randomly distributed particles will have a higher probability to settle in a minimum with small $\Omega_s$ compared to another minimum (when the energies and degeneracy factors are similar).

 \begin{table}[h]
 \caption{Energy difference between metastable states and the ground state (ground state energy given in italic numbers) that were used to compute the partition function for $N=27$ and $\kappa=0.6$. Also shown is the relative statistical weight $\tilde{n}_s=n_s/n_{1}$ and the statistical weight due to the eigenfrequencies $\tilde{w}_s=\left(\Omega_{1}/\Omega_s\right)^f$ compared to the ground state.\label{tab:statesN27}}
 \begin{ruledtabular}
 \begin{tabular}{ccrrr}
 state $s$& configuration & $\Delta E_s/N$ & $\tilde{n}_s$ & $\tilde{w}_s$  \\
 \hline
 1 & (24,3) & $\textsl{4.732856(4)}$ & $1$ & $1$\\
 2 & (23,4) & $0.001622(1)$ & $6$ & $0.24$\\
 3 & (23,4) & $0.001870(5)$ & $6$ & $0.67$\\
 4 & (25,2) & $0.004993(0)$ & $3/25$ & $14$\\
 5 & (25,2) & $0.004997(3)$ & $3/25$ & $3.3$\\
 \end{tabular}
 \end{ruledtabular}
 \end{table}

Because the harmonic approximation only describes a minimum's local neighborhood we mention that this could overestimate the weight of states with broad minima and low escape paths~\cite{baletto2005}, which are not taken into account in this approximation. This could be improved by changing the limits for the position integration in Eq.~(\ref{eqn:classical}) according to the potential barrier height and the temperature. This was done for 2D clusters in~\cite{schweigert1995} but requires knowledge of the barrier heights. This is not essential for the present analysis. Finally, we note that the value of $\tilde{w}_s$ is sensitive to numerical errors in the computation of the eigenvalues of the Hessians since the mean eigenfrequency is a product of $3N-3$ single values. In the present results we estimate the error not to exceed $5\,\%$ which is sufficient for our analysis.

\begin{table}
 \caption{Same as Table~\ref{tab:statesN27} for $N=31$ and $\kappa=0.8$.\label{tab:statesN31}}
 \begin{ruledtabular}
 \begin{tabular}{ccrrr}
 state $s$& configuration & $\Delta E_s^0/N$ & $\tilde{n}_s$ & $\tilde{w}_s$  \\
 \hline
 1 & (27,4) & $\textsl{4.397858(8)}$ & $1$ & $1$\\
 2 & (27,4) & $0.000008(7)$ & $1$ & $0.82$\\
 3 & (27,4) & $0.000035(8)$ & $1$ & $1.7$\\
 4 & (26,5) & $0.001810(1)$ & $27/5$ & $0.84$\\
 5 & (26,5) & $0.001850(9)$ & $27/5$ & $1.4$\\
 6 & (26,5) & $0.002000(0)$ & $27/5$ & $5.3$\\
 7 & (26,5) & $0.002091(6)$ & $27/5$ & $9.7$\\
 8 & (25,6) & $0.003583(7)$ & $117/5$ & $1.4$\\
 9 & (25,6) & $0.003586(7)$ & $117/5$ & $1.1$\\ 
 \end{tabular}
 \end{ruledtabular}
 \end{table}
\begin{table}
 \caption{Same as Table~\ref{tab:statesN27} for $N=40$ and $\kappa=0$ (Coulomb interaction).\label{tab:statesN40}}
 \begin{ruledtabular}
 \begin{tabular}{ccrrr}
 state $s$& configuration & $\Delta E_s^0/N$ & $\tilde{n}_s$ & $\tilde{w}_s$\\
\hline
1 & (34,6) & $\textsl{12.150162(9)}$ & $1$ & $1$\\
2 & (33,7) & $0.001143(4)$ & $34/7$ & $2.3$\\
3 & (33,7) & $0.001190(3)$ & $34/7$ & $2.8$\\
4 & (33,7) & $0.001236(9)$ & $34/7$ & $8.3$\\
5 & (32,8) & $0.001862(8)$ & $561/28$ & $13$\\
6 & (32,8) & $0.001863(1)$ & $561/28$ & $3.5$\\
7 & (32,8) & $0.003482(4)$ & $561/28$ & $6.7$\\
8 & (35,5) & $0.004201(7)$ & $6/35$ & $5.2$\\
9 & (35,5) & $0.004392(7)$ & $6/35$ & $32$\\
 \end{tabular}
 \end{ruledtabular}
 \end{table}

 \begin{figure}
 \centering
{\includegraphics*[width=0.42\textwidth]{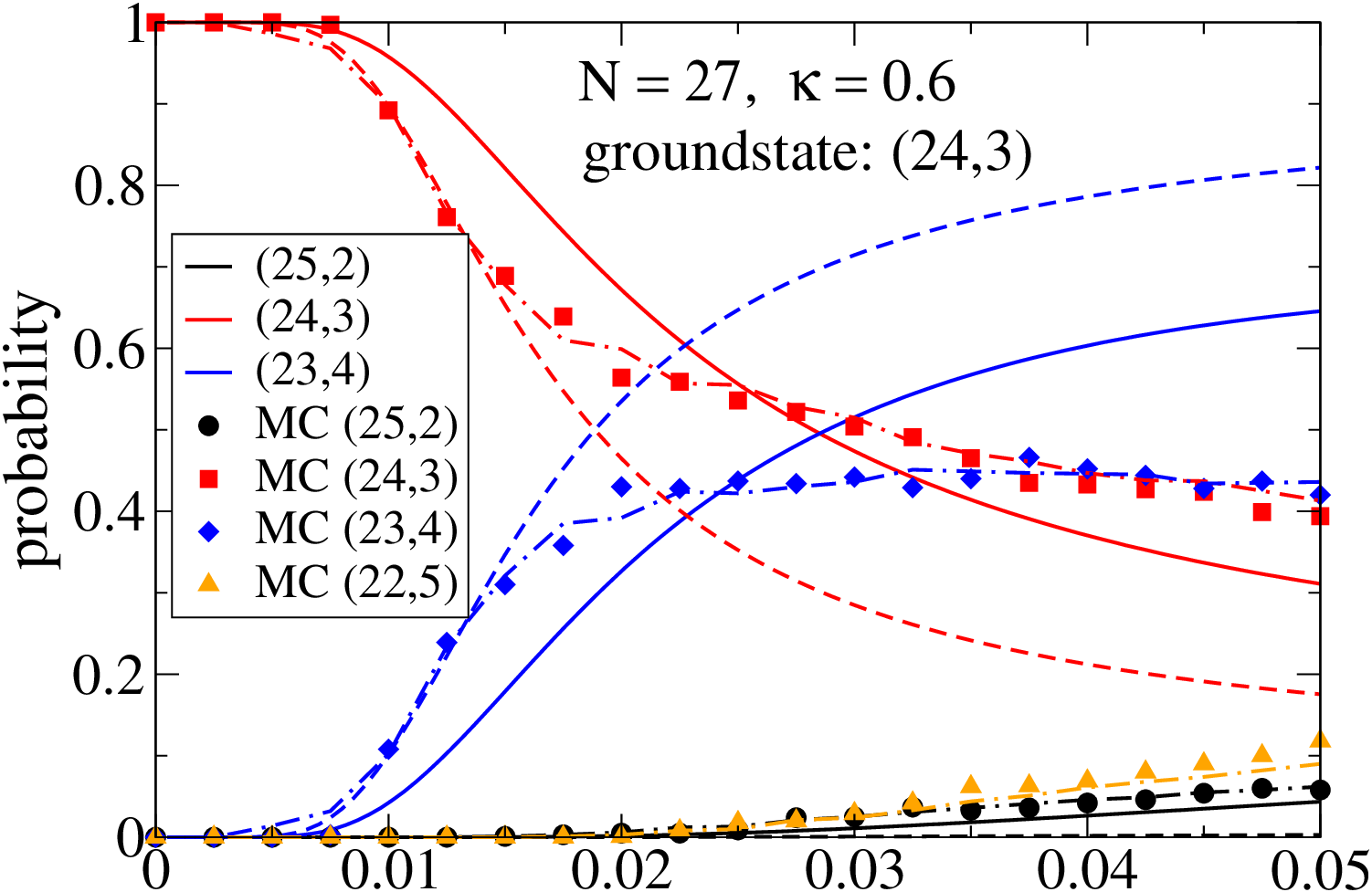}}\vspace{0.13cm}
{\includegraphics*[width=0.42\textwidth]{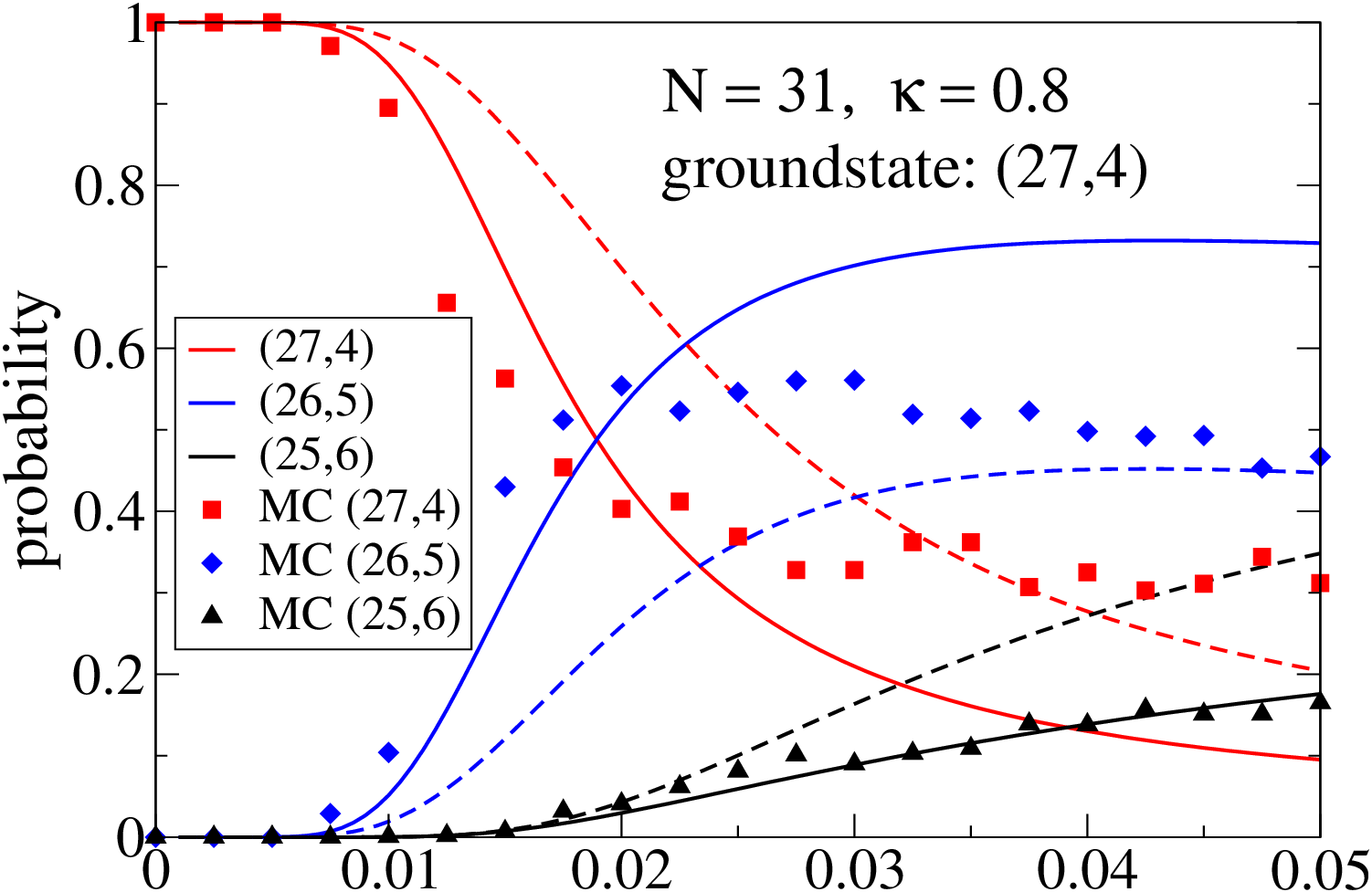}}\vspace{0.13cm}
{\includegraphics*[width=0.42\textwidth]{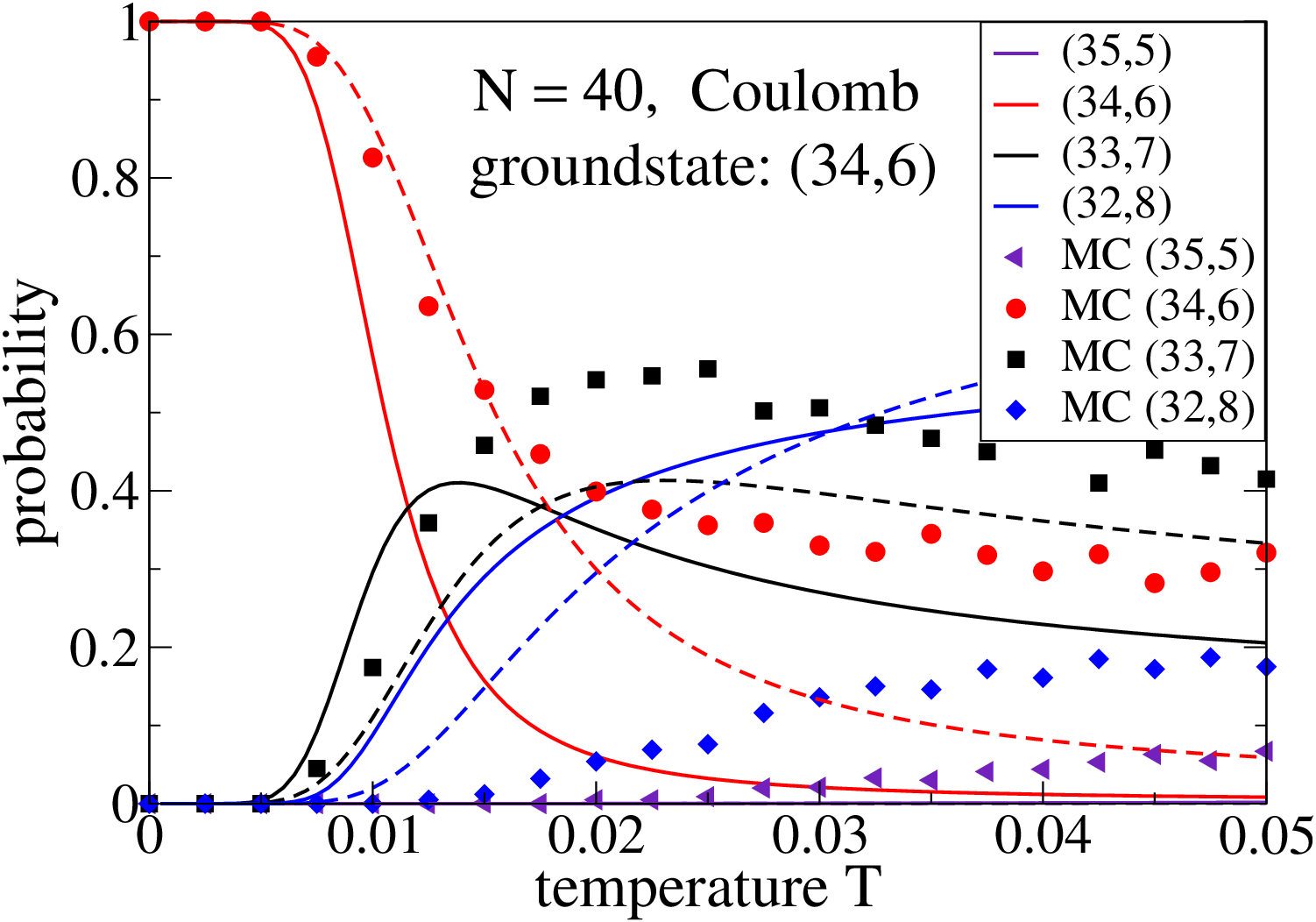}}
 \caption{(Color online) Analytical theory compared to MC-results. The solid lines show the probabilities as obtained from Eq.~(\ref{eqn:Probability}). The dashed lines neglect the statistical weight factor caused by the eigenfrequencies, i.e. here $\Omega_s\equiv 1$ for all states. For $N=27$ the dashed/dotted lines indicate the results of Langevin dynamics simulations. Analytical results for the configuration (22,5) are not available.}{\label{fig:MC_Analyt}}
 \end{figure}

\subsection{{\label{ssec:model_mc}}Analytical results and comparison with Monte Carlo simulations}
Let us now come to the results of the analytical model and 
compare them to Monte-Carlo simulations which were explained above in Section~\ref{ssec:MC}. The MC results have first principle character, in particular, they are not restricted to the harmonic approximation and fully include all anharmonic corrections. For $N=27$ we additionally verified the MC results by a Langevin dynamics simulation using the SLO algorithm of~\cite{Langevin}. Here the probabilities were obtained in an equilibrium calculation with a simulation time $t=10^5\,\omega_0^{-1}$ by determining the configurations at fixed time intervals.

Results for three representative examples are shown in Fig.~\ref{fig:MC_Analyt}.
We chose $N=27,\;\kappa=0.6$ and $N=31,\;\kappa=0.8$ since these will turn out to be close to the situation in the dusty plasma experiments, see. Sec.~\ref{sec:comp}. As a third example we present data for $N=40$ with Coulomb interaction. The input parameters of the analytical model, i.e. details on the (metastable) states are summarized in Tables~\ref{tab:statesN27}-\ref{tab:statesN40}.
In Fig.~\ref{fig:MC_Analyt} we plot the occurrence probabilities as a function of temperature. This allows us to specifically study the effect of the depth of the potential energy minimum $E_s$. The latter effect should be dominant at low temperature, leading to a relatively high probability of the ground state. In contrast, this effect should become less important at high temperature where the degeneracy factors and the eigenfrequency ratio should play a decisive role for the probabilities.
This general trend is indeed observed in all three cases.

For $N=27$, top part of Fig.~\ref{fig:MC_Analyt}, the effects of the degeneracy factor and the mean eigenfrequencies act in opposite directions. While the state with $4$ particles on the inner shell gains statistical weight by having a high degeneracy, this effect is almost compensated by narrow minima and, consequently, a low $\tilde{w}_s$, cf.~Tab.~\ref{tab:statesN27}. Therefore, this state 
achieves comparable probability with the ground state (27,4) only at high temperature, $T \ge 0.03$ (in the MC simulation this is observed only for $T \ge 0.045$).
For the configuration (25,2) the opposite is true. Here, the degeneracy is low and the minima broad, but due to its high energy this configuration has a nonvanishing probability only for high temperatures, $T \ge 0.03$. We did not find a stable state with configuration (22,5).

The situation for $N=31$, central part of Fig.~\ref{fig:MC_Analyt}, is different. Here all metastable states have a higher degeneracy factor than the ground state configuration. In addition all states further gain statistical weight because of broad minima, except for state $s=4$, cf.~Tab.~\ref{tab:statesN31}. Thus one should expect that metastable states have a high probability even at low temperatures. This is indeed observed in the model and the MC simulations already below $T=0.02$.

In the third case, $N=40$, bottom part of Fig.~\ref{fig:MC_Analyt}, we generally see the same trend. 
The metastable state (33,7) has a high degeneracy and frequency factor, cf.~Tab.~\ref{tab:statesN31}, 
and thus it becomes more probable than the ground state already for $T\ge 0.01$ ($0.015$ in the MC simulations). 

Let us now compare the analytical and MC results more in detail. Good agreement is found for $N=31$ up to $T\sim 0.02$, cf. full lines and symbols. For $N=27$ we find good agreement between MC and the analytical theory for $T<0.012$ but only if the effect of the eigenfrequencies is neglected, cf. dashed lines. With eigenfrequencies included the theory shows deviations for low temperatures but better agreement for higher temperatures. For the cluster with $40$ particles we observe moderate agreement for the configurations (34,6) and (33,7) up to $T=0.015$ whereas the deviations from MC for the remaining two configurations are rather large. This overall agreement is quite satisfactory keeping in mind that the melting temperature of these clusters is typically below $T=0.015$ \cite{vova2006,Apolinario07,Jens2008}.

The reason for these discrepancies are due to the limitations of our simple harmonic model [the good agreement between the completely independent MC and Langevin MD results for $N=27$, cf. top part of Fig.~\ref{fig:MC_Analyt}, confirms the reliability of the simulations].
Since the discrepancies are growing with temperature, the main reason is probably the neglect of anharmonic effects. In some cases, when the barriers of the potential energy surface are low, these effects might already occur at low temperatures. Changing the limits of allowed particle motion in the integration of Eq.~(\ref{eqn:classical}) may help to reduce the deviations. 
A further reason for deviations from MC results could be an insufficient number of stationary states being taken into account. It is not clear if all stationary states have been found (they were pre-computed with MD simulations) and used in the partition function. To ensure a high probability we performed more than $10^4$ independent runs. For example, for the cluster with 40 particles we observe 9 states, but it was difficult to identify the states with 5 particles on the inner shell because they were found only a few times and were energetically close. The larger number of states given in~\cite{Apolinario} also suggests that we missed a few. Nevertheless, the effect originating from these states should give only a small statistical contribution to the probabilities.

\section{Comparison with dusty plasma experiments}\label{sec:comp}
To compare with  experiments  on metastable states in spherical dusty plasma crystals (Yukawa balls) we first need to establish the relation of our system of units to the experimental parameters. We use the temperature unit $k_B T_0=E_0=(\alpha Q^4/2)^{1/3}$ [in SI units $E_0=(\alpha Q^4/32\pi^2\epsilon_0^2)^{1/3}$] which depends on the trap parameter $\alpha=m\omega_0^2$ and the dust charge. Since the charge is not known very accurately the errors could be rather large. With $Z=2000\,e$ and $\alpha=5.2\times 10^{-11}\,\text{kg}\,\text{s}^{-2}$ given in~\cite{block2008}, room temperature ($300\,\text{K}$) corresponds to $T_{room}\approx 0.0015$. Also, the experimental screening parameter is known only approximately. From previous comparisons with simulations \cite{bonitz2006} it is expected to be in the range of $0.5 < \kappa < 1$. Reference~\cite{block2008}
reported measurements on the probability of metastable states for two clusters with $N=27$ and $N=31$  which we now use for comparison with the MD and MC simulations and the analytical model.

\subsection{{\label{ssec:md_exp}}MD results vs. experiment}
\begin{figure}
 {\includegraphics*[width=0.45\textwidth]{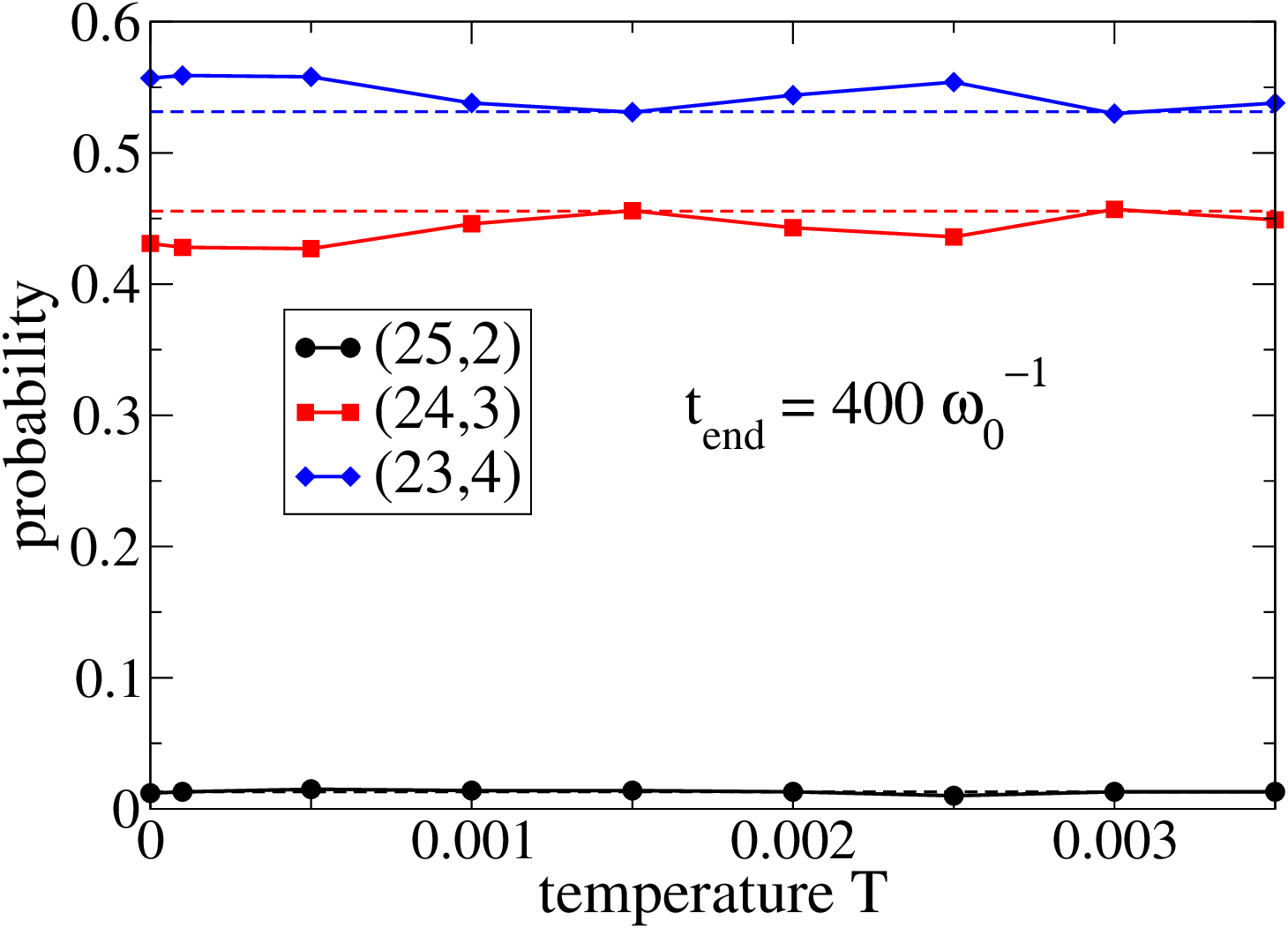}}
 \caption{Langevin dynamics simulation for $N=27$, $\kappa=0.6$ and $\nu=3.2$. Horizontal lines indicate results of Section~\ref{ssec:screening}, Fig.~\ref{fig:kappa}.}{\label{fig:Tfinite}}
 \end{figure}

We start with the molecular dynamics simulations since they model a situation which is closest to the experiment. In contrast to the experiment which is performed at room temperature, our simulations correspond to a Langevin dynamics simulation at $T=0$ (the system is cooled to almost zero kinetic energy). We have verified the influence of the final temperature by performing additional Langevin simulations for the cluster with 27 particles and $\kappa=0.6$ with temperatures up to $T=0.0035$ (Fig.~\ref{fig:Tfinite}) which is more than twice the experimental temperature. Apart from a finite temperature the simulations were done in the same way as explained in Section~\ref{sec:model}, but with a predefined simulation time. For high temperatures one has to pay attention to the time after which the configuration is determined since then transitions between states can easily occur. This can be seen in Fig.~\ref{fig:MC_Analyt} where for $T>0.01$ metastable states have a nonvanishing probability. In our Langevin simulations we used a simulation time of $t_{end}=400\,\omega_0^{-1}$, which corresponds to $t_{end}\approx 10\,\text{s}$ for a dust particle mass of $m=3.3\times 10^{-14}\,\text{kg}$. We find no systematic deviation from the results at zero temperature. The slight deviations for the configurations (23,4) and (24,3) are probably due to the insufficiently long simulation time with the same explanation as given at the end of Section~\ref{ssec:screening}. We thus conclude that for the present analysis an MD simulation without fluctuations and cooling towards zero temperature is adequate.

Our data for comparison with the experimental results are shown in Figs.~\ref{fig:friction} and \ref{fig:kappa}. The friction parameter in the experiments is expected to be in the range $\nu=3\dots 6$ \cite{block2008}. This means the system is overdamped and any value above $\nu=2$ will not change the results significantly, cf. Fig.~\ref{fig:friction}.  So in Fig.~\ref{fig:kappa} we used a value of $3.2$.
The MD simulations agree well with the experiment in the case of screening parameters in the range $0.6<\kappa<0.8$ (for $N=31$) and $0.4<\kappa<0.6$ ($N=27$), for details cf. Table~\ref{tab:comp}. The lower screening parameter in the latter case is a consequence of the lower plasma density in the experiment, compared to the conditions under which the cluster with 31 particles was produced. This was also found in the MD simulations performed in~\cite{block2008}. The present simulations, being much more extensive, confirm these results. We may conclude that this comparison allows to determine the screening parameter in the experiment.

\subsection{{\label{ssec:model_exp}}Analytical and MC results vs. experiment}
A comparison of the analytical model and the MC simulations with the experiment is disappointing. 
From Fig.~\ref{fig:MC_Analyt} it is evident that at room temperature the ground states have always a probability of almost $100\,\%$ which is in striking contrast to the experiment and the MD results. 
This is not surprising since the dust comprises a dissipative system and the clusters are created under nonequilibrium conditions. In contrast, both Monte Carlo and the model are based on the canonical partition function and assume thermodynamic equilibrium. Thus, at first sight, there seems to be no way to explain the experiment with our analytical model or with Monte Carlo methods. However, this is not true. As we will show below, there is a way to apply equilibrium methods to the problem of metastable states.

\subsection{{\label{ssec:dynamics}}Time scales of the cluster dynamics}
Let us have a closer look at the nonequilibrium dynamics of the cluster during the cooling process. It is particularly interesting to analyze on what time scales the different relaxation processes occur. 
In a weakly coupled plasma there are three main time scales, e.g. \cite{bonitz-book,semkat99}: first, the buildup of binary correlations which occurs for times shorter than the correlation time $\tau_{cor}$. Second, the relaxation of the velocity distribution towards local equilibrium due to collisions, for $\tau_{cor}\le t \le t_{rel}$ (kinetic phase) and third, hydrodynamic relaxation, $t_{rel}\le t\le t_{hyd}$. 
This behavior has so far not been analyzed for the strongly correlated Yukawa clusters. 

\begin{figure}
 {\includegraphics*[width=0.48\textwidth]{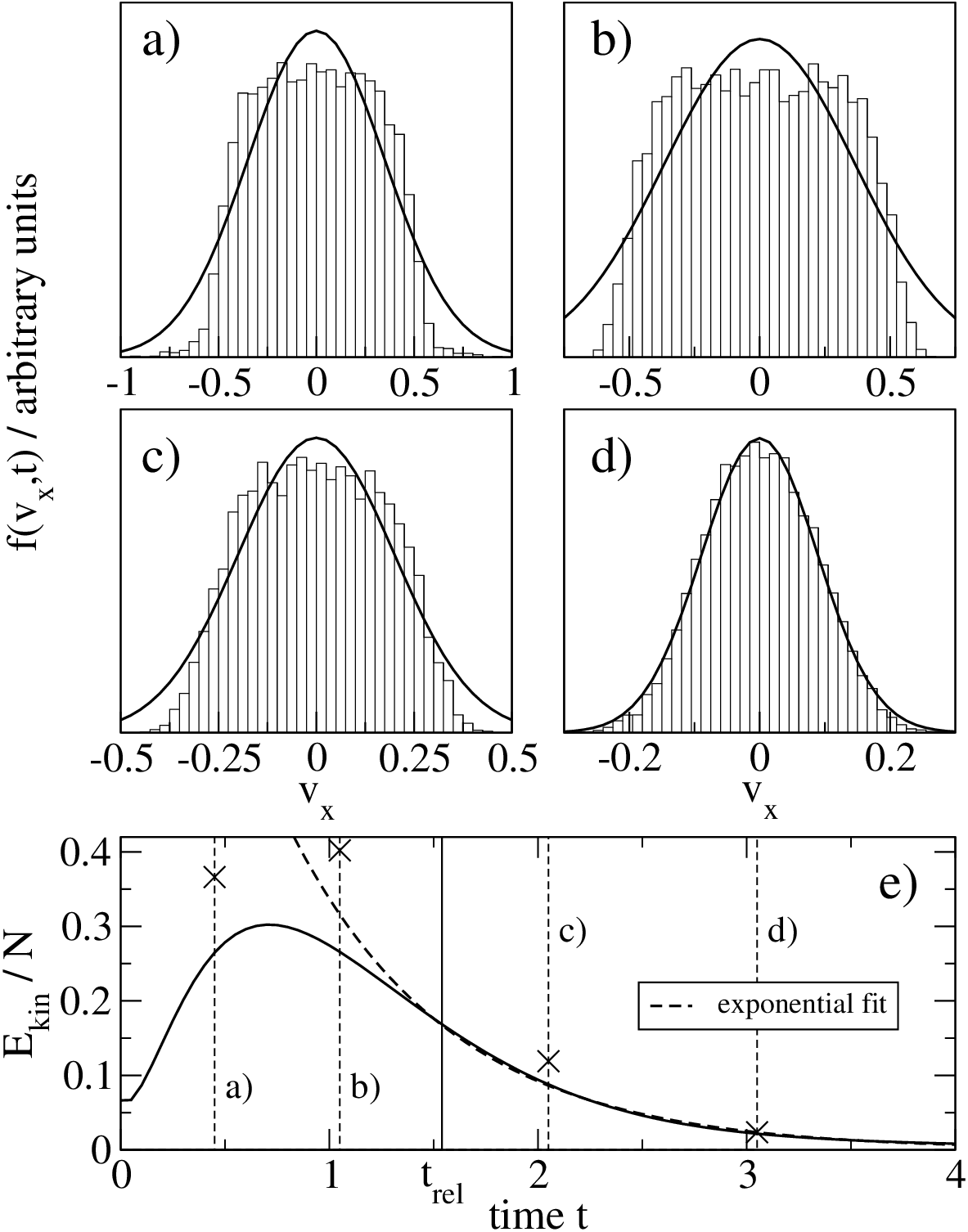}}
 \caption{a) - d) Velocity distribution function $f(v_x,t)$ for different times [as indicated in e)] for $N=27$, $\kappa=0.6$, $\nu=3.2$. Solid lines show the best Maxwellian fit. e) Averaged kinetic energy as a function of time. Crosses denote averaged kinetic energy obtained from best fit using the equipartition theorem.}{\label{fig:distribution}}
 \end{figure}

To get first insight, the quantities of central interest are the kinetic energy and the velocity distribution function $f(v,t)$ of the cluster particles. These quantities are easily computed in our nonequilibrium MD simulations of the cooling process, as explained in Section~\ref{sec:MDresults}. To obtain the velocity distribution we performed 420 runs with different randomly chosen initial conditions and collected the data for each time step. The results for the kinetic energy evolution and for $f(v_x,t)$ are shown in Fig.~\ref{fig:distribution} (the other velocity components show the same behavior). We observe three main relaxation stages:  
\begin{enumerate}
 \item for $t\le 0.5$, a rapid heating is observed which is due to acceleration and build up of binary correlations in the initially random (uncorrelated) particle system. This is typical for any rapid change of the interparticle forces,  and proceeds on scales of the order of the correlation time, e.g. \cite{bonitz96,haberland01,gericke03}. 
 \item for $0.5\le t\le 1.3$, the kinetic energy increase saturates and cooling starts. This means, correlation build up is finished and dissipation due to neutral gas friction dominates the behavior. 
 \item for $t>1.3$, the mean kinetic energy decreases approximately exponentially, i.e. $\langle E_{\rm kin}\rangle(t) \propto e^{-2\gamma t}$ where the decay constant is found to be $\gamma\approx 0.65\approx \nu/5$.
\end{enumerate}
The behavior on the third stage resembles a single (Brownian) particle in a dissipative medium where $\gamma$ is the velocity relaxation rate corresponding to a relaxation time of $t_{rel}=\gamma^{-1}=1.54$. In case of Brownian particles, the velocity distribution rapidly relaxes towards a Maxwellian for $t\le t_{rel}$. The velocity distributions for the present system are shown 
for four different times in Fig.~\ref{fig:distribution}, parts a)-d). The solid curves indicate the best fit to a Maxwellian, the obtained ``temperatures'' are shown in Fig.~\ref{fig:distribution}~e) by the crosses. The evolution towards a Maxwellian is evident which is established around $t=2.5$.

This allows us to conclude that, after an initial stage (phases 1 and 2), the cluster has reached an equilibrium velocity distribution and the subsequent cooling process ultimately leading to freezing into a spherical Yukawa crystal is well described by local thermodynamic equilibrium: the time-dependent velocity distribution is given by
$f(v,t)\sim \exp\{-\frac{mv^2}{2k_BT(t)}\}$ with $k_B T(t)=2\langle E_{kin}\rangle(t)/3$.
Thus, the system evolves from one equilibrium state to another which differ only by temperature.

\subsection{{\label{ssec:lte}}Application of Equilibrium Theories to the probability of metastable states of Yukawa balls}
Based on the results of Subsection~\ref{ssec:dynamics}, we expect that equilibrium methods such as Monte Carlo or our analytical model are applicable to the third relaxation stage. Thereby one has to use the equilibrium result for the current temperature $T(t)$. Using temperature dependent results such as in Fig.~\ref{fig:MC_Analyt}, allows one to reconstruct the time-dependence of various quantities from the known dynamics of the kinetic energy: $T(t)=T(t_{rel}) e^{-2\gamma (t-t_{rel})}$.

Now, the key point is that this local (time-dependent) Maxwellian is established long before the particles start to feel the potential energy $U$ of the trap and of the pair interaction. For example, at $t\approx t_{rel}$, the temperature is around $0.15$ which is about a factor $100$ higher than room temperature and one order of magnitude higher than the freezing point. In case of very rapid cooling beyond the freezing point the particles will settle (with a certain probability) in the stationary state ``s'' and will not have time to escape it since further cooling removes the necessary kinetic energy (i.e. the escape probability will be low). This means that the decision about what stationary state the system will reach is made at a time when the system temperature is close to the melting temperature. 

Using this idea we compute the probability of metastable states from Monte Carlo for two temperatures $T=0.02$ and $T=0.04$, cf. Fig.~\ref{fig:MC_Analyt} (at the higher temperature, due to intershell transitions, shell configurations can be identified only with an error of about $8\%$). We also calculate the probability at $T=0.02$ within the analytical model. Finally we consider the high-temperature limit which is obtained by neglecting, in the probability ratios, the Boltzmann factor. The corresponding results are presented in Table~\ref{tab:comp}. The overall agreement with the experiment is much better than the results for room temperature which confirms the correctness of the above arguments. Evidently, the Boltzmann factor is crucial and cannot be neglected, cf. last lines in Table~\ref{tab:comp}. The best results are observed for temperatures around $T=0.04$ which is about two to three times higher than the melting temperature where the system is in the moderately coupled liquid state.
This shows that it is indeed possible to predict, at least qualitatively, the probabilities of metastable states in dissipative nonequilibrium Yukawa crystals within equilibrium models and simulations. This is possible in the overdamped limit as is the case in dusty plasmas.

 \begin{table}
 \caption{Comparison of experimental results for $N=27$ and $N=31$ with MD and MC simulations (MC results are for the two temperatures $T=0.02$ and $T=0.04$). Also shown are the results of the  analytical model (``AM'') for $T=0.02$ and with the Boltzmann factor being neglected ($T\rightarrow \infty$). For $N=27$ ($N=31$) the simulation results are shown for $\kappa=0.6$ ($\kappa=0.8$).}
 \begin{ruledtabular}\label{tab:comp}
 \begin{tabular}{cllll}
 $N=27$ & $P(24,3)$ & $P(23,4)$ & $P(25,2)$\\
 \hline
Experiment & $0.46\pm 0.14$ & $0.46\pm 0.14$ & $0.08\pm 0.06$\\
MD & $0.46$ & $0.53$ & $0.01$\\
MC(0.02) & $0.56$ & $0.43$ & $0.01$\\
MC(0.04) & $0.43$ & $0.45$ & $0.04$\\
AM(0.02) & $0.67$ & $0.33$ & $0.00$\\
AM($\infty$) & $0.12$ & $0.64$ & $0.24$\\
\hline \hline 
 $N=31$ & $P(27,4)$ & $P(26,5)$ & $P(25,6)$\\
 \hline
Experiment & $0.35\pm 0.10$ & $0.62\pm 0.13$ & $0.03\pm 0.03$\\
MD & $0.30$ & $0.59$ & $0.11$\\
MC(0.02) & $0.40$ & $0.55$ & $0.04$\\
MC(0.04) & $0.33$ & $0.50$ & $0.14$\\
AM(0.02) & $0.44$ & $0.53$ & $0.03$\\
AM($\infty$) & $0.02$ & $0.60$ & $0.38$\\
 \end{tabular}
 \end{ruledtabular}
 \end{table}

\section{Discussion}
In summary we have presented simulation results for Yukawa balls with three different numbers of particles and a broad range of screening parameters and damping coefficients. It was shown by extensive molecular dynamics and Langevin dynamics simulations that the cooling speed (damping coefficient) strongly affects the occurrence probabilities of metastable states even if the interaction and the confinement remain the same. This is similar to the liquid solid transition in macroscopic systems where rapid cooling may give rise to a glass-like disordered solid rather than a crystal with lower total energy. The same scenario is also observed in the present finite crystals.
While slow cooling leads predominantly to the lowest energy state, strong damping gives rise to an increased probability of metastable states. These states may have an up to five times higher probability than the ground state, which is fully consistent with the recent observation of metastable states in dusty plasma experiments~\cite{block2008}. These metastable states are not an artefact of an imperfect experiment or due to fluctuations of experimental parameters, but are an intrinsic property of finite Yukawa balls.

Furthermore we showed that screening strongly alters the results compared to Coulomb interaction. Generally increased screening leads to a higher probability of states with more particles on inner shells due to the shorter interaction range. An analytical theory for the ground state density profile of a confined one-component Yukawa plasma~\cite{Christian2006, Christian2007} also showed that decreasing the screening length (increasing $\kappa$) leads to a higher particle density in the center of the trap, which would correspond to a higher population of inner shells in our case.

We presented an analytical model based on the canonical partition function and the harmonic approximation for the total potential energy. This model allowed for a physically intuitive explanation of the observed high probabilities of metastable configurations. The Boltzmann factor (which always favors the ground state relative to higher lying states), competes with two 
factors that favor metastable states: the degeneracy factor [favoring states with more particles on the inner shell(s)] and the local curvature of the potential minimum. Low curvature (low eigenfrequency) corresponds to a broad minimum and a large phase space volume attracting particles.
Among all normal modes the dominant effect is due to the energetically lowest modes. 
The thermodynamic results from Monte-Carlo simulations and the analytical theory are in reasonable agreement with each other, at low temperatures, as expected. For higher temperatures anharmonic effects such as barrier heights will be equally important.

It was shown that in thermodynamic equilibrium the abundances of metastables are much lower than observed in the dusty plasma experiments at the same temperature. The reason is that, in equilibrium, the particles are given infinitely long time to escape a local potential minimum and they always will visit the ground state more frequently than any metastable state. In contrast, in the limit of strong damping the particles are being trapped in the first minimum they visit. Thus the decision about the final stationary state is being made early during the cooling process, when the temperature is of the order of two to three times the melting temperature. Therefore, equilibrium theories without dissipation may be successfully applied to strongly correlated and strongly damped nonequilibrium systems. A systematic derivation from a time-dependent theory is still lacking and will be presented in a forthcoming paper.

\begin{acknowledgements}
 We acknowledge stimulating discussions with J.W. Dufty. This work is supported by the Deutsche Forschungsgemeinschaft via SFB-TR 24 and by the U.S. Department of Energy award DE-FG02-07ER54946.
\end{acknowledgements}

\bibliography{probability3}

\begin{thebibliography}{25}
\expandafter\ifx\csname natexlab\endcsname\relax\def\natexlab#1{#1}\fi
\expandafter\ifx\csname bibnamefont\endcsname\relax
  \def\bibnamefont#1{#1}\fi
\expandafter\ifx\csname bibfnamefont\endcsname\relax
  \def\bibfnamefont#1{#1}\fi
\expandafter\ifx\csname citenamefont\endcsname\relax
  \def\citenamefont#1{#1}\fi
\expandafter\ifx\csname url\endcsname\relax
  \def\url#1{\texttt{#1}}\fi
\expandafter\ifx\csname urlprefix\endcsname\relax\def\urlprefix{URL }\fi
\providecommand{\bibinfo}[2]{#2}
\providecommand{\eprint}[2][]{\url{#2}}

\bibitem[{\citenamefont{Wineland et~al.}(1987)\citenamefont{Wineland,
  Bergquist, Itano, Bollinger, and Manney}}]{Wineland1987}
\bibinfo{author}{\bibfnamefont{D.~J.} \bibnamefont{Wineland}},
  \bibinfo{author}{\bibfnamefont{J.~C.} \bibnamefont{Bergquist}},
  \bibinfo{author}{\bibfnamefont{W.~M.} \bibnamefont{Itano}},
  \bibinfo{author}{\bibfnamefont{J.~J.} \bibnamefont{Bollinger}},
  \bibnamefont{and} \bibinfo{author}{\bibfnamefont{C.~H.}
  \bibnamefont{Manney}}, \bibinfo{journal}{Phys. Rev. Lett.}
  \textbf{\bibinfo{volume}{59}}, \bibinfo{pages}{2935} (\bibinfo{year}{1987}).

\bibitem[{\citenamefont{Drewsen et~al.}(1998)\citenamefont{Drewsen, Brodersen,
  Hornek\ae{}r, Hangst, and Schiffer}}]{drewsen1998}
\bibinfo{author}{\bibfnamefont{M.}~\bibnamefont{Drewsen}},
  \bibinfo{author}{\bibfnamefont{C.}~\bibnamefont{Brodersen}},
  \bibinfo{author}{\bibfnamefont{L.}~\bibnamefont{Hornek\ae{}r}},
  \bibinfo{author}{\bibfnamefont{J.~S.} \bibnamefont{Hangst}},
  \bibnamefont{and} \bibinfo{author}{\bibfnamefont{J.~P.}
  \bibnamefont{Schiffer}}, \bibinfo{journal}{Phys. Rev. Lett.}
  \textbf{\bibinfo{volume}{81}}, \bibinfo{pages}{2878} (\bibinfo{year}{1998}).

\bibitem[{\citenamefont{Filinov et~al.}(2001)\citenamefont{Filinov, Bonitz, and
  Lozovik}}]{Alex2001}
\bibinfo{author}{\bibfnamefont{A.~V.} \bibnamefont{Filinov}},
  \bibinfo{author}{\bibfnamefont{M.}~\bibnamefont{Bonitz}}, \bibnamefont{and}
  \bibinfo{author}{\bibfnamefont{Y.~E.} \bibnamefont{Lozovik}},
  \bibinfo{journal}{Phys. Rev. Lett.} \textbf{\bibinfo{volume}{86}},
  \bibinfo{pages}{3851} (\bibinfo{year}{2001}).

\bibitem[{\citenamefont{Arp et~al.}(2004)\citenamefont{Arp, Block, Piel, and
  Melzer}}]{Arp2004}
\bibinfo{author}{\bibfnamefont{O.}~\bibnamefont{Arp}},
  \bibinfo{author}{\bibfnamefont{D.}~\bibnamefont{Block}},
  \bibinfo{author}{\bibfnamefont{A.}~\bibnamefont{Piel}}, \bibnamefont{and}
  \bibinfo{author}{\bibfnamefont{A.}~\bibnamefont{Melzer}},
  \bibinfo{journal}{Phys. Rev. Lett.} \textbf{\bibinfo{volume}{93}},
  \bibinfo{pages}{165004} (\bibinfo{year}{2004}).

\bibitem[{\citenamefont{Bonitz et~al.}(2008)\citenamefont{Bonitz, Ludwig,
  Baumgartner, Henning, Filinov, Block, Arp, Piel, K\"ading, Ivanov
  et~al.}}]{bonitz2008pop}
\bibinfo{author}{\bibfnamefont{M.}~\bibnamefont{Bonitz}},
  \bibinfo{author}{\bibfnamefont{P.}~\bibnamefont{Ludwig}},
  \bibinfo{author}{\bibfnamefont{H.}~\bibnamefont{Baumgartner}},
  \bibinfo{author}{\bibfnamefont{C.}~\bibnamefont{Henning}},
  \bibinfo{author}{\bibfnamefont{A.}~\bibnamefont{Filinov}},
  \bibinfo{author}{\bibfnamefont{D.}~\bibnamefont{Block}},
  \bibinfo{author}{\bibfnamefont{O.}~\bibnamefont{Arp}},
  \bibinfo{author}{\bibfnamefont{A.}~\bibnamefont{Piel}},
  \bibinfo{author}{\bibfnamefont{S.}~\bibnamefont{K\"ading}},
  \bibinfo{author}{\bibfnamefont{Y.}~\bibnamefont{Ivanov}},
  \bibnamefont{et~al.}, \bibinfo{journal}{Physics of Plasmas}
  \textbf{\bibinfo{volume}{15}}, \bibinfo{eid}{055704} (\bibinfo{year}{2008}).

\bibitem[{\citenamefont{Ludwig et~al.}(2005)\citenamefont{Ludwig, Kosse, and
  Bonitz}}]{Patrick2005}
\bibinfo{author}{\bibfnamefont{P.}~\bibnamefont{Ludwig}},
  \bibinfo{author}{\bibfnamefont{S.}~\bibnamefont{Kosse}}, \bibnamefont{and}
  \bibinfo{author}{\bibfnamefont{M.}~\bibnamefont{Bonitz}},
  \bibinfo{journal}{Phys. Rev. E} \textbf{\bibinfo{volume}{71}},
  \bibinfo{eid}{046403} (\bibinfo{year}{2005}).

\bibitem[{\citenamefont{Arp et~al.}(2005)\citenamefont{Arp, Block, Bonitz,
  Fehske, Golubnychiy, Kosse, Ludwig, Melzer, and Piel}}]{Arp2005}
\bibinfo{author}{\bibfnamefont{O.}~\bibnamefont{Arp}},
  \bibinfo{author}{\bibfnamefont{D.}~\bibnamefont{Block}},
  \bibinfo{author}{\bibfnamefont{M.}~\bibnamefont{Bonitz}},
  \bibinfo{author}{\bibfnamefont{H.}~\bibnamefont{Fehske}},
  \bibinfo{author}{\bibfnamefont{V.}~\bibnamefont{Golubnychiy}},
  \bibinfo{author}{\bibfnamefont{S.}~\bibnamefont{Kosse}},
  \bibinfo{author}{\bibfnamefont{P.}~\bibnamefont{Ludwig}},
  \bibinfo{author}{\bibfnamefont{A.}~\bibnamefont{Melzer}}, \bibnamefont{and}
  \bibinfo{author}{\bibfnamefont{A.}~\bibnamefont{Piel}},
  \bibinfo{journal}{Journal of Physics: Conference Series}
  \textbf{\bibinfo{volume}{11}}, \bibinfo{pages}{234} (\bibinfo{year}{2005}).

\bibitem[{\citenamefont{Baumgartner et~al.}(2008)\citenamefont{Baumgartner,
  Golubnychiy, Asmus, Ludwig, and Bonitz}}]{Henning}
\bibinfo{author}{\bibfnamefont{H.}~\bibnamefont{Baumgartner}},
  \bibinfo{author}{\bibfnamefont{V.}~\bibnamefont{Golubnychiy}},
  \bibinfo{author}{\bibfnamefont{D.}~\bibnamefont{Asmus}},
  \bibinfo{author}{\bibfnamefont{P.}~\bibnamefont{Ludwig}}, \bibnamefont{and}
  \bibinfo{author}{\bibfnamefont{M.}~\bibnamefont{Bonitz}},
  \bibinfo{journal}{submitted for publication}  (\bibinfo{year}{2008}).

\bibitem[{\citenamefont{Apolinario et~al.}(2007)\citenamefont{Apolinario,
  Partoens, and Peeters}}]{Apolinario}
\bibinfo{author}{\bibfnamefont{S.~W.~S.} \bibnamefont{Apolinario}},
  \bibinfo{author}{\bibfnamefont{B.}~\bibnamefont{Partoens}}, \bibnamefont{and}
  \bibinfo{author}{\bibfnamefont{F.~M.} \bibnamefont{Peeters}},
  \bibinfo{journal}{New Journal of Physics} \textbf{\bibinfo{volume}{9}},
  \bibinfo{pages}{283} (\bibinfo{year}{2007}).

\bibitem[{\citenamefont{Bonitz et~al.}(2006)\citenamefont{Bonitz, Block, Arp,
  Golubnychiy, Baumgartner, Ludwig, Piel, and Filinov}}]{bonitz2006}
\bibinfo{author}{\bibfnamefont{M.}~\bibnamefont{Bonitz}},
  \bibinfo{author}{\bibfnamefont{D.}~\bibnamefont{Block}},
  \bibinfo{author}{\bibfnamefont{O.}~\bibnamefont{Arp}},
  \bibinfo{author}{\bibfnamefont{V.}~\bibnamefont{Golubnychiy}},
  \bibinfo{author}{\bibfnamefont{H.}~\bibnamefont{Baumgartner}},
  \bibinfo{author}{\bibfnamefont{P.}~\bibnamefont{Ludwig}},
  \bibinfo{author}{\bibfnamefont{A.}~\bibnamefont{Piel}}, \bibnamefont{and}
  \bibinfo{author}{\bibfnamefont{A.}~\bibnamefont{Filinov}},
  \bibinfo{journal}{Phys Rev. Lett.} \textbf{\bibinfo{volume}{96}},
  \bibinfo{eid}{075001} (\bibinfo{year}{2006}).

\bibitem[{\citenamefont{Block et~al.}(2008)\citenamefont{Block, K\"{a}ding,
  Melzer, Piel, Baumgartner, and Bonitz}}]{block2008}
\bibinfo{author}{\bibfnamefont{D.}~\bibnamefont{Block}},
  \bibinfo{author}{\bibfnamefont{S.}~\bibnamefont{K\"{a}ding}},
  \bibinfo{author}{\bibfnamefont{A.}~\bibnamefont{Melzer}},
  \bibinfo{author}{\bibfnamefont{A.}~\bibnamefont{Piel}},
  \bibinfo{author}{\bibfnamefont{H.}~\bibnamefont{Baumgartner}},
  \bibnamefont{and} \bibinfo{author}{\bibfnamefont{M.}~\bibnamefont{Bonitz}},
  \bibinfo{journal}{Physics of Plasmas} \textbf{\bibinfo{volume}{15}},
  \bibinfo{eid}{040701} (\bibinfo{year}{2008}).

\bibitem[{\citenamefont{Baletto and Ferrando}(2005)}]{baletto2005}
\bibinfo{author}{\bibfnamefont{F.}~\bibnamefont{Baletto}} \bibnamefont{and}
  \bibinfo{author}{\bibfnamefont{R.}~\bibnamefont{Ferrando}},
  \bibinfo{journal}{Reviews of Modern Physics} \textbf{\bibinfo{volume}{77}},
  \bibinfo{eid}{371} (\bibinfo{year}{2005}).

\bibitem[{\citenamefont{Filinov and Bonitz}(2006)}]{computational}
\bibinfo{author}{\bibfnamefont{A.}~\bibnamefont{Filinov}} \bibnamefont{and}
  \bibinfo{author}{\bibfnamefont{M.}~\bibnamefont{Bonitz}}, in
  \emph{\bibinfo{booktitle}{Introduction to Computational Methods in Many-Body
  Physics}}, edited by \bibinfo{editor}{\bibfnamefont{M.}~\bibnamefont{Bonitz}}
  \bibnamefont{and} \bibinfo{editor}{\bibfnamefont{D.}~\bibnamefont{Semkat}}
  (\bibinfo{publisher}{Rinton Press}, \bibinfo{address}{Princeton},
  \bibinfo{year}{2006}).

\bibitem[{\citenamefont{Schweigert and Peeters}(1995)}]{schweigert1995}
\bibinfo{author}{\bibfnamefont{V.~A.} \bibnamefont{Schweigert}}
  \bibnamefont{and} \bibinfo{author}{\bibfnamefont{F.~M.}
  \bibnamefont{Peeters}}, \bibinfo{journal}{Phys. Rev. B}
  \textbf{\bibinfo{volume}{51}}, \bibinfo{pages}{7700} (\bibinfo{year}{1995}).

\bibitem[{\citenamefont{Mannella}(2004)}]{Langevin}
\bibinfo{author}{\bibfnamefont{R.}~\bibnamefont{Mannella}},
  \bibinfo{journal}{Phys. Rev. E} \textbf{\bibinfo{volume}{69}},
  \bibinfo{pages}{041107} (\bibinfo{year}{2004}).

\bibitem[{\citenamefont{Golubnychiy et~al.}(2006)\citenamefont{Golubnychiy,
  Baumgartner, Bonitz, Filinov, and Fehske}}]{vova2006}
\bibinfo{author}{\bibfnamefont{V.}~\bibnamefont{Golubnychiy}},
  \bibinfo{author}{\bibfnamefont{H.}~\bibnamefont{Baumgartner}},
  \bibinfo{author}{\bibfnamefont{M.}~\bibnamefont{Bonitz}},
  \bibinfo{author}{\bibfnamefont{A.}~\bibnamefont{Filinov}}, \bibnamefont{and}
  \bibinfo{author}{\bibfnamefont{H.}~\bibnamefont{Fehske}},
  \bibinfo{journal}{J. Phys. A: Math. Gen.} \textbf{\bibinfo{volume}{39}},
  \bibinfo{pages}{4527} (\bibinfo{year}{2006}).

\bibitem[{\citenamefont{Apolinario and Peeters}(2007)}]{Apolinario07}
\bibinfo{author}{\bibfnamefont{S.~W.~S.} \bibnamefont{Apolinario}}
  \bibnamefont{and} \bibinfo{author}{\bibfnamefont{F.~M.}
  \bibnamefont{Peeters}}, \bibinfo{journal}{Phys. Rev. E}
  \textbf{\bibinfo{volume}{76}}, \bibinfo{pages}{031107}
  (\bibinfo{year}{2007}).

\bibitem[{\citenamefont{B\"oning et~al.}(2008)\citenamefont{B\"oning, Filinov,
  Ludwig, Baumgartner, Bonitz, and Lozovik}}]{Jens2008}
\bibinfo{author}{\bibfnamefont{J.}~\bibnamefont{B\"oning}},
  \bibinfo{author}{\bibfnamefont{A.}~\bibnamefont{Filinov}},
  \bibinfo{author}{\bibfnamefont{P.}~\bibnamefont{Ludwig}},
  \bibinfo{author}{\bibfnamefont{H.}~\bibnamefont{Baumgartner}},
  \bibinfo{author}{\bibfnamefont{M.}~\bibnamefont{Bonitz}}, \bibnamefont{and}
  \bibinfo{author}{\bibfnamefont{Y.}~\bibnamefont{Lozovik}},
  \bibinfo{journal}{Phys. Rev. Lett.} \textbf{\bibinfo{volume}{100}},
  \bibinfo{pages}{113401} (\bibinfo{year}{2008}).

\bibitem[{\citenamefont{Bonitz}(1998)}]{bonitz-book}
\bibinfo{author}{\bibfnamefont{M.}~\bibnamefont{Bonitz}}, in
  \emph{\bibinfo{booktitle}{Quantum Kinetic Theory}}
  (\bibinfo{publisher}{Teubner}, \bibinfo{address}{Stuttgart, Leipzig},
  \bibinfo{year}{1998}).

\bibitem[{\citenamefont{Semkat et~al.}(1999)\citenamefont{Semkat, Kremp, and
  Bonitz}}]{semkat99}
\bibinfo{author}{\bibfnamefont{D.}~\bibnamefont{Semkat}},
  \bibinfo{author}{\bibfnamefont{D.}~\bibnamefont{Kremp}}, \bibnamefont{and}
  \bibinfo{author}{\bibfnamefont{M.}~\bibnamefont{Bonitz}},
  \bibinfo{journal}{Phys. Rev. E} \textbf{\bibinfo{volume}{59}},
  \bibinfo{pages}{1557} (\bibinfo{year}{1999}).

\bibitem[{\citenamefont{Bonitz and Kremp}(1996)}]{bonitz96}
\bibinfo{author}{\bibfnamefont{M.}~\bibnamefont{Bonitz}} \bibnamefont{and}
  \bibinfo{author}{\bibfnamefont{D.}~\bibnamefont{Kremp}},
  \bibinfo{journal}{Phys. Lett. A} \textbf{\bibinfo{volume}{212}},
  \bibinfo{pages}{83} (\bibinfo{year}{1996}).

\bibitem[{\citenamefont{Haberland et~al.}(2001)\citenamefont{Haberland, Bonitz,
  and Kremp}}]{haberland01}
\bibinfo{author}{\bibfnamefont{H.}~\bibnamefont{Haberland}},
  \bibinfo{author}{\bibfnamefont{M.}~\bibnamefont{Bonitz}}, \bibnamefont{and}
  \bibinfo{author}{\bibfnamefont{D.}~\bibnamefont{Kremp}},
  \bibinfo{journal}{Phys. Rev. E} \textbf{\bibinfo{volume}{64}},
  \bibinfo{pages}{026405} (\bibinfo{year}{2001}).

\bibitem[{\citenamefont{Gericke et~al.}(2003)\citenamefont{Gericke, Murillo,
  Semkat, Bonitz, and Kremp}}]{gericke03}
\bibinfo{author}{\bibfnamefont{D.}~\bibnamefont{Gericke}},
  \bibinfo{author}{\bibfnamefont{M.}~\bibnamefont{Murillo}},
  \bibinfo{author}{\bibfnamefont{D.}~\bibnamefont{Semkat}},
  \bibinfo{author}{\bibfnamefont{M.}~\bibnamefont{Bonitz}}, \bibnamefont{and}
  \bibinfo{author}{\bibfnamefont{D.}~\bibnamefont{Kremp}}, \bibinfo{journal}{J.
  Phys.A: Math. Gen.} \textbf{\bibinfo{volume}{36}}, \bibinfo{pages}{6087}
  (\bibinfo{year}{2003}).

\bibitem[{\citenamefont{Henning et~al.}(2006)\citenamefont{Henning,
  Baumgartner, Piel, Ludwig, Golubnichiy, Bonitz, and Block}}]{Christian2006}
\bibinfo{author}{\bibfnamefont{C.}~\bibnamefont{Henning}},
  \bibinfo{author}{\bibfnamefont{H.}~\bibnamefont{Baumgartner}},
  \bibinfo{author}{\bibfnamefont{A.}~\bibnamefont{Piel}},
  \bibinfo{author}{\bibfnamefont{P.}~\bibnamefont{Ludwig}},
  \bibinfo{author}{\bibfnamefont{V.}~\bibnamefont{Golubnichiy}},
  \bibinfo{author}{\bibfnamefont{M.}~\bibnamefont{Bonitz}}, \bibnamefont{and}
  \bibinfo{author}{\bibfnamefont{D.}~\bibnamefont{Block}},
  \bibinfo{journal}{Phys. Rev. E} \textbf{\bibinfo{volume}{74}},
  \bibinfo{eid}{056403} (\bibinfo{year}{2006}).

\bibitem[{\citenamefont{Henning et~al.}(2007)\citenamefont{Henning, Ludwig,
  Filinov, Piel, and Bonitz}}]{Christian2007}
\bibinfo{author}{\bibfnamefont{C.}~\bibnamefont{Henning}},
  \bibinfo{author}{\bibfnamefont{P.}~\bibnamefont{Ludwig}},
  \bibinfo{author}{\bibfnamefont{A.}~\bibnamefont{Filinov}},
  \bibinfo{author}{\bibfnamefont{A.}~\bibnamefont{Piel}}, \bibnamefont{and}
  \bibinfo{author}{\bibfnamefont{M.}~\bibnamefont{Bonitz}},
  \bibinfo{journal}{Phys. Rev. E} \textbf{\bibinfo{volume}{76}},
  \bibinfo{eid}{036404} (\bibinfo{year}{2007}).

\end{thebibliography}

\end{document}